\documentclass[11pt]{article}
\usepackage{xcolor}
\usepackage{cite}
\usepackage{multirow}
\usepackage{enumerate}
\usepackage[scale=0.8]{geometry}

\usepackage{graphicx}
\usepackage[subrefformat=parens,labelformat=parens]{subfig}
\usepackage{hyperref}

\usepackage{amsfonts,amsmath}
\usepackage{multirow,multicol}
\usepackage{bbm}
\usepackage{xspace}
\usepackage{slashed}

\usepackage[sort&compress, english]{cleveref}

\newcommand{\bC}{\ensuremath{\mathbbm{C}}\xspace}
\newcommand{\bP}{\ensuremath{\mathbbm{P}}\xspace}

\newcommand{\FS}{\ensuremath{\text{FS}}\xspace}
\newcommand{\CY}{\ensuremath{\text{CY}}\xspace}

\newcommand{\cL}{\mathcal{L}}

\newcommand{\gpred}{g_{\text{pr}}}%{g_{\text{pred}}}

\newcommand{\gCY}{{\ensuremath{g_\CY}\xspace}}
\newcommand{\gFSX}{g_{\text{FS}}}

\newcommand{\vCY}{\text{Vol}_{\text{CY}}}
\newcommand{\vFS}{\text{Vol}_{\text{FS}}}
\newcommand{\vO}{\text{Vol}_{\Omega}}

\newcommand{\kp}{{\mathcal K}}

\renewcommand{\Re}{{\rm Re}\;} 
\renewcommand{\Im}{{\rm Im }\;} 

\numberwithin{equation}{section}

\begin{document}
\begin{center}
{
\LARGE {\bf Numerical Metrics for Complete Intersection and Kreuzer-Skarke Calabi-Yau Manifolds}\\[12pt]
\vspace{1cm}
\normalsize
{\bf{Magdalena Larfors$^{a,b}$}\footnote{magdalena.larfors@physics.uu.se}},
{\bf{Andre Lukas$^{c,}$}\footnote{andre.lukas@physics.ox.ac.uk}},
{\bf{Fabian Ruehle$^{d,e,}$}\footnote{f.ruehle@northeastern.edu}},
{\bf{Robin Schneider$^{b,}$}\footnote{robin.schneider@physics.uu.se}}
\bigskip}\\[0pt]
\vspace{0.23cm}
${}^a$ {\it 
Department of Mathematical Sciences, Durham University,\\
Upper Mountjoy Campus, Stockton Rd, Durham DH1 3LE, UK
}\\[2ex]
${}^b$ {\it 
Department of Physics and Astronomy, Uppsala University\\
SE-751 20 Uppsala, Sweden
}\\[2ex]
${}^c$ {\it 
Rudolf Peierls Centre for Theoretical Physics, University of Oxford\\
Parks Road, Oxford OX1 3PU, UK
}\\[2ex]
${}^d$ {\it 
Department of Physics \& Department of Mathematics, Northeastern University\\
360 Huntington Avenue, Boston MA 02115 United States
}\\[2ex]
${}^e$ {\it 
The NSF AI Institute for Artificial Intelligence and Fundamental Interactions
}
\end{center}
\vspace{0.5cm}

\begin{abstract}\noindent
We introduce neural networks to compute numerical Ricci-flat CY metrics for complete intersection and Kreuzer-Skarke Calabi-Yau manifolds at any point in K\"ahler and complex structure moduli space, and introduce the package \texttt{cymetric} which provides computation realizations of these techniques. In particular, we develop and computationally realize methods for point-sampling on these manifolds. The training for the neural networks is carried out subject to a custom loss function. The K\"ahler class is fixed by adding to the loss a component which enforces the slopes of certain line bundles to match with topological computations. Our methods are applied to various manifolds, including the quintic manifold, the bi-cubic manifold and a Kreuzer-Skarke manifold with Picard number two. We show that volumes and line bundle slopes can be reliably computed from the resulting Ricci-flat metrics. We also apply our results to compute an approximate Hermitian-Yang-Mills connection on a specific line bundle on the bi-cubic. 
\end{abstract}

\setcounter{footnote}{0}
\setcounter{tocdepth}{2}
\clearpage
\tableofcontents

%%%%%%%%%%%%%%%%%%%%%%%%%%%%%%%%%%%%%%%%%%%%%%%%%%%%%%%%%%%%%%%%%%%%%%%%%%%%
\section{Introduction}
\label{sec:introduction}

Calabi-Yau (CY) manifolds arise naturally in the context of string theory~\cite{Candelas:1985en}. Since the 80's mathematicians and physicists alike have invested a considerable amount of effort in constructing and studying these spaces. By now there are different popular constructions, such as complete intersection hypersurfaces in products of complex projective spaces~\cite{Green:1987cr,Candelas:1987kf} (CICYs) or more generally in toric varieties~\cite{Kreuzer:2000xy}. CY three-folds constructed as hypersurfaces in toric four-folds are described by about half a billion reflexive polytopes (giving rise to many more topologically inequivalent manifolds) which we refer to as the Kreuzer-Skarke (KS) list. Much is known about their topological properties~\cite{Hubsch:1992nu}, such as their Chern classes, intersection numbers or Hodge numbers. However, objects related to their differential geometry are much less accessible. Crucially, the unique Ricci-flat metric on CY manifolds, established by theorems of Calabi and Yau~\cite{Calabi+2015+78+89,Yau:1978cfy}, is not known in analytic form on any compact CY three-folds, although there is recent progress for K3~\cite{Kachru:2020tat,Kachru:2018van}.

The Ricci-flat CY metric is of great importance to string theorists, as it plays a role in many computations connecting string theory with observable physics. For example, the CY metric is required to compute the physical Yukawa couplings from string theory ~\cite{green2012superstring} or it can be used to study the physics which arises along geodesics in moduli space, in order to probe the swampland distance conjecture~\cite{Ashmore:2021qdf}. There are also mathematical problems which benefit from knowledge of the CY metric, including the SYZ-conjecture~\cite{Strominger:1996it}, which aims to formalize mirror symmetry~\cite{hori2003mirror}. Given the importance of the Ricci-flat CY metric and the lack of analytic expressions it is only natural that several numerical approximation schemes have been advanced.

The first work in this direction relies on an algorithm set up by Donaldson~\cite{donaldson2005numerical}. This algorithm starts with an approximation for the K\"ahler potential underlying the Ricci-flat metric which is of the form $K = \ln( s^i h_{ij} \bar{s}^j)$, where $s^i$ is a basis of sections for the $k^{\rm th}$ power, $L^k$, of a line bundle $L\rightarrow X$ on the CY manifold $X$ and $h_{ij}$ is a hermitian matrix. For each power $k$ an optimal $K$ for a balanced metric can be found by a fix point method which computes the matrix $h_{ij}$. In the limit $k\rightarrow \infty$ these balanced metrics converge to the Ricci-flat CY metric with K\"ahler class proportional to $c_1(L)$. However, the scaling of this method with $k$ makes it unsuitable for cases without a high degree of symmetry or with many K\"ahler parameters. In particular the algorithm requires a factorial increase in training data for a sub-linear increase in accuracy~\cite{Douglas:2006hz}. Another practical problem is the link between the K\"ahler class of the Ricci-flat metric and the lattice-valued quantity $c_1(L)$ which makes it difficult to determine the metric as a function of the K\"ahler parameters. Nevertheless this method has been used to construct metrics on the Fermat Quintic and other selected examples~\cite{Douglas:2006rr,Braun:2007sn,Braun:2008jp,Douglas:2006hz,Afkhami-Jeddi:2021qkf} and it has been generalized to compute the solution to the Hermitian Yang-Mills equation for vector bundles~\cite{Anderson:2010ke,Anderson:2011ed}.

Scaling of the computation can be improved by a method which relies on energy functionals~\cite{Headrick:2009jz}. This method also uses a polynomial Ansatz for the (K\"ahler potential of the) metric and optimizes polynomial coefficients by minimizing the deviation from the Monge-Ampere equation. In this way solutions with exponential accuracy are found. The code\footnote{\url{https://people.brandeis.edu/~headrick/Mathematica/index.html}} published alongside~\cite{Headrick:2009jz} for CY hypersurfaces of Fermat type has been used in more recent studies~\cite{Cui:2019uhy} to investigate the curvature near conifold points and to accurately predict the K\"ahler potential~\cite{Ashmore:2021ohf}.

The rise of neural networks as universal function approximators has renewed the interest in finding numerical approximations of the Ricci-flat CY metric. They have been introduced into the string community in 2017~\cite{He:2017aed,Ruehle:2017mzq,Krefl:2017yox,Carifio:2017bov} (see~\cite{Ruehle:2020jrk} for a review) and by now several directions to use techniques from machine learning for CY metrics have emerged. Decision trees and curve fitting have been used to bootstrap from low order Donaldson algorithms to more accurate solutions~\cite{Ashmore:2019wzb}. Holomorphic neural networks are shown to learn much better and more efficient approximations of the K\"ahler potential~\cite{Douglas:2020hpv}. Finally, the metric has been learned directly with neural networks in~\cite{Anderson:2020hux,Jejjala:2020wcc,Larfors:2021pbb}. These new numerical methods allowed for direct application to test the swampland distance conjecture~\cite{Ashmore:2021qdf} and construct solutions to the Hermitian Yang-Mills equation of line bundles~\cite{Ashmore:2021rlc}.

While there has been progress on methods to compute the Ricci-flat CY metric, there are also still considerable gaps, particularly in view of providing systematic tools, for example in the context of string model building. Computations have been carried out for select manifolds but a tool which covers large classes of CY manifolds is still missing. This requires an understanding of some theoretical problems, including methods for point sampling on more general CY manifolds. A critical issue for applications is the ability to compute the CY metric at a given point in moduli space. This tends to be straightforward for the complex structure moduli which are usually fixed by the equations defining the CY manifold. For a single K\"ahler modulus, $h^{1,1}(X)=1$, such as for the quintic, things are simple as every K\"ahler class can be obtained by a simple re-scaling of the metric. For $h^{1,1}(X)>1$ this is no longer the case and care has to be taken in order to fix the desired K\"ahler class. While this is possible in principle by Donaldson's algorithm, computations are limited by inefficiency and the limitation to integral K\"ahler classes. On the other hand, neural network approaches which solely rely on minimizing the Ricci curvature or deviations form the Monge-Ampere equation produce a metric with an unspecified K\"ahler class.

The point of the present paper is to address and resolve these issues and provide both the theoretical background and practical methods for machine-learning Ricci-flat CY metrics at a prescribed point in moduli space, for any manifold from both the CICY and KS lists. This is extending and detailing recent work by the authors in Ref.~\cite{Larfors:2021pbb}. In particular, we explain how the problem of point sampling on CY manifolds can be solved for both CICY and KS manifolds. We fix the K\"ahler class by matching line bundles slopes, which can be computed using topological data only, to their numeric evaluation using the approximate \CY metric. For a given K\"ahler class $J$, we consider a basis $L_i$ of line bundles and determine the slopes $\mu_J(L_i)$ using a topological formula for the slope. During training, a numerical version of these slopes, $\mu^{\rm num}_J(L_i)$, is computed by numerical integration and a loss proportional to $\sum_i||\mu_J(L_i)-\mu^{\rm num}_J(L_i)||_n$ is added to the loss function. As we will demonstrate, this does indeed lead to an approximately Ricci-flat metric with the prescribed K\"ahler class $J$.

These methods are realized in the \texttt{cymetric} package which provides machine learning tools to find the Ricci-flat CY metric for both CICY and KS manifolds. The package is written in Python and Mathematica~\cite{Mathematica}. The point generation relies on routines from \texttt{NumPy}~\cite{harris2020array} and \texttt{SciPy}~\cite{mckinney-proc-scipy-2010}. For KS manifolds we employ routines implemented in SageMath~\cite{sagemath} and Mathematica~\cite{Mathematica}. The neural networks are implemented and optimized with \texttt{TensorFlow}~\cite{tensorflow2015-whitepaper}, which has efficient tracing for several nested gradient operations using automatic differentiation. The code is available at:
\begin{center}
    \url{https://github.com/pythoncymetric/cymetric}
\end{center}

Open source software has led to rapid progress in machine learning. Our goal is to achieve similarly remarkable results by open sourcing our code and we welcome contributions. There are several other open source packages developed by the string theory community. For example the toric geometry routines of SageMath~\cite{toricroutines} are an integral part of the workflow for many geometers. They rely on the \texttt{PALP} package~\cite{Kreuzer:2002uu} and are a necessary component in our toric point generator. The functionality of these routines have recently been expanded by the \texttt{CYTools} package~\cite{cytools}. Other open source packages include Stringvacua~\cite{Gray:2008zs} for studying potentials from string theory and \texttt{cohomcalc}~\cite{cohomCalg:Implementation} and \texttt{pyCICY}~\cite{CICYtoolkit} for studying line bundle cohomologies over CY manifolds.

The outline of this paper is as follows. In Section~\ref{sec:background} we will introduce the necessary mathematical background for the computation of the Ricci-flat metric, including the construction of the required K\"ahler forms and a discussion of the point sampling method.  Section~\ref{sec:hym} reviews some basic mathematics related to Hermitian Yang-Mills (HYM) connections on line bundles and the numerical method for computing such connections. A brief review of neural networks and an overview of our network architecture follows in Section~\ref{sec:compbackground}. Results for the quintic and bi-cubic CY manifolds as well as a Picard number two manifold from the KS list are presented in Section~\ref{sec:results} and we conclude in Section~\ref{sec:conclusions}. Some useful details on toric geometry are collection in Appendix~\ref{app:ToricGeometryIntro}.

%%%%%%%%%%%%%%%%%%%%%%%%%%%%%%%%%%%%%%%%%%%%%%%%%%%%%%%%%%%%%%%%%%%%%%%%%%%%
\section{Mathematical background}
\label{sec:background}

In this section we introduce the mathematical background necessary for sampling points and calculating the CY metric. For our procedure we require three pieces of data for the CY manifold: Its holomorphic top form $\Omega$, a K\"ahler form $J$, and the affine patches plus their transition functions. We collect here only the important formulas and provide an introduction to the necessary concepts in toric geometry in Appendix~\ref{app:ToricGeometryIntro}. We adopt a convention where we add subscripts \FS and \CY to quantities related to the ambient Fubini-Study metric (such as the metric itself or its associated K\"ahler potential) and the Ricci-flat Calabi-Yau metric, respectively. We do not distinguish between ambient space quantities and their restriction to not clutter the notation further and it should be clear from the context which one we are referring to. Our formulae will be written for CY three-folds, but they generalize straight-forwardly to CY $n$-folds\footnote{The code supports CY manifolds up to complex dimension 6.}. We denote real CY coordinates on a patch by $y^m$, where $m =1,\ldots ,6$ and their complex counterparts by
\begin{equation}
 z^a=\frac{1}{\sqrt{2}}(y^a+iy^{a+3})\;,\qquad \bar{z}^{\bar{a}}=\frac{1}{\sqrt{2}}(y^{\bar{a}}-iy^{\bar{a}+3})\; ,
\end{equation} 
where $a,b=1,2,3$ and $\bar{a},\bar{b} =1,2,3$. 

\subsection{K\"ahler forms and volumes}

To start, let us recall that the Calabi-Yau theorem~\cite{Calabi+2015+78+89,Yau:1978cfy} states that any K\"ahler $n$-fold $X$ with $c_1(TX)=0$ and given K\"ahler form $J'$ admits a unique Ricci-flat metric $g_\CY$ (simple called CY metric, in the following) whose associated K\"ahler form $J_\CY$ (where $J_{\CY,a\bar{b}}=ig_{\CY,a\bar{b}}$) is in the same cohomology class as $J'$. For three-folds, the K\"ahler form $J_\CY$ can be determined from the Monge-Amp\`ere equation
\begin{align}
		\label{eq:MA}
		J_\CY \wedge J_\CY \wedge J_\CY = \kappa \; \Omega \wedge \bar{\Omega}  \qquad \text{ with } \qquad  J_\CY = J' + \partial \bar{\partial} \phi		
\end{align}
where $\phi$ is a real function on $X$ and $\kappa$ is a complex constant (which may depend on moduli). The K\"ahler form $J'$ can be any reference K\"ahler form in the given K\"ahler class but for our purposes we will take this to be the obvious K\"ahler form induced from the toric ambient space, that is the (generalized) Fubini-Study K\"ahler form denoted by $J'=J_\FS$. It is usually written as a linear combination
\begin{equation}\label{JFSdef}
 J_\FS=t^\alpha J_{\alpha}
\end{equation}
relative to a basis  $J_{\alpha}$ of $(1,1)$ forms (where $\alpha = 1,\ldots ,h^{1,1}(X)$), also obtained by restriction from the ambient space. Here $t^\alpha$ are the K\"ahler parameters. Associated to these K\"ahler forms are volume forms
\begin{equation}\label{eq:volforms}
 \text{d} \vCY=\frac{1}{3!}J_\CY^3\;,\qquad \text{d} \vFS=\frac{1}{3!}J_\FS^3
\end{equation}
and a typical problem is to compute integrals of the type
\begin{equation}
 \int_X \text{d} \vCY\, f=\int_X d^{6}y\sqrt{g_\CY}\,f\;,\qquad \int_X \text{d} \vFS\, f=\int_Xd^{6}y\sqrt{g_\FS}\, f\; ,
\end{equation}
for some function $f$ on $X$. We will see later how numerical results for $J_\CY$ (and, indeed, $J_\FS$) can be used the evaluate such integrals numerically. For the constant function $f=1$ this computes the CY volume
\begin{equation}\label{CYvol}
 %{\rm Vol}_t(X)
  {\rm Vol}_{\CY}=\int_X \text{d} \vCY=\int_X\text{d} \vFS=\frac{1}{3!}d_{\alpha\beta\gamma}t^\alpha t^\beta t^\gamma
\end{equation}
which coincides for the CY and FS measure (provided, as we assume, $J_\CY$ and $J_\FS$ are in the same cohomology class) and which can be computed from the topological formula on the RHS of Eq.~\eqref{CYvol}. Here
\begin{equation}
 d_{\alpha\beta\gamma}=\int_XJ_{\alpha}\wedge J_{\beta}\wedge J_{\gamma}\; .
\end{equation}
are the intersection numbers which are known or can be computed for CICY and KS manifolds (our package does this automatically in the background).
\subsection{Line bundle slopes}\label{sec:linebundles}
Another interesting set of quantities which can be used for the same purpose (and we will, in fact, use it to enforce the correct K\"ahler class during training) is the slope of line bundles. The line bundle $L\rightarrow X$ with first Chern class $c_1(L)=[k^\alpha J_{\alpha}]$ is denoted by ${\cal O}_X(k)$ and its slope $\mu_t({\cal O}_X(k))$ is defined by
\begin{equation}\label{eq:slopeizec}
 \mu_t({\cal O}_X(k))=\int_X J_\CY^2\wedge c_1({\cal O}_X(k))=\int_X J_\FS^2\wedge c_1({\cal O}_X(k))=d_{\alpha\beta\gamma}\,t^\alpha t^\beta k^\gamma\; .
\end{equation}
Note, since the intersection numbers are known, this can be computed from the K\"ahler parameters $t^\alpha$ and the line bundle integers $k^\alpha$. On the other hand, introducing the curvature $F_\FS$ and corresponding sources $\rho_\FS$ and  $\rho_\CY$  on ${\cal O}_X(k)$ by
\begin{equation}\label{eq:rhodef}
 F_\FS=2\pi i k^\alpha J_{\alpha}\;,\quad \rho_\FS=\frac{1}{2}g_\FS^{a\bar{b}}F_{\FS,a\bar{b}}\;,\quad \rho_\CY=\frac{1}{2}g_\CY^{a\bar{b}}F_{\FS,a\bar{b}}
\end{equation} 
the slope can also be computed from
\begin{equation}\label{eq:slope}
 \mu_t({\cal O}_X(k))=\frac{2}{\pi}\int_X\text{d}\vCY\;\rho_\CY=\frac{2}{\pi}\int_X\text{d} \vFS\; \rho_\FS\; .
\end{equation}
The numerical version of this equation will be used to compute line bundle slopes during training and compute the slope loss, in order to enforce the correct K\"ahler class.
 
\subsection{Constructing the holomorphic top form}\label{sec:topform}
For our purposes we need to construct the holomorphic top form $\Omega$ explicitly and we follow a method described, for example, in Ref.~\cite{Cox:2011aaa}. For a KS manifold, defined as a hypersurface in a toric four-fold ${\cal A}$ there are always four affine ambient space coordinates $z^\mu$, where $\mu=1,\ldots ,4$, on a given patch, explicitly given in Eq.~\eqref{eq:affinecoordinates}. Then we can write the holomorphic $(3,0)$ form as
\begin{align}
\Omega=\frac{dz_1\wedge dz_2 \wedge dz_3}{\partial p/\partial z_4}\,,
\end{align}
where the coordinate $z_4$ in the denominator should be replaced by a solution $z_4=z_4(z_a)$, where $a=1,2,3$, of the defining equation $p(z_\mu)=0$. More details on the underlying toric geometry can be found in Appendix~\ref{app:ToricGeometryIntro}.

For the case of CICYs, defined in an ambient space ${\cal A}=\mathbb{P}^{n_1}\times\cdots\times\mathbb{P}^{n_h}$ with total dimension $d=\sum_{\alpha=1}^h n_\alpha$, we have affine coordinates $z^\mu$, where $\mu=1,\ldots ,d$ on each patch. From these we can choose three coordinates, $z^a$, where $a=1,2,3$, to parameterize the CY patch while the remaining $K=d-3$ coordinates $z^q$ can be written as functions $z^q=z^q(z^a)$ by using the $K$ defining polynomials $p_r$ of the CICY. The holomorphic $(3,0)$-form can then be written in terms of the determinant of the $K\times K$ matrix $\partial p_r/\partial z^q$, and is explicitly given by
\begin{align}
\Omega=\frac{dz_1\wedge dz_2 \wedge dz_3}{\det(\partial p_r/\partial z^q)}\,,
\end{align}
where the coordinates $z^q$ in the denominator have to be replaced by the solutions $z^q=z^q(z^a)$. 

Given $\Omega$, it is useful to introduce the associated volume form
\begin{equation}\label{dVCYdVO}
 \text{d}\vO=\Omega\wedge\bar{\Omega}\qquad\Rightarrow\qquad \text{d}\vCY=\frac{\kappa}{3!}\text{d}\vO
\end{equation}
which, by virtue of the Monge-Ampere equation~\eqref{eq:MA} and Eq.~\eqref{eq:volforms}, is proportional to the CY volume form. This means, while the Ricci-flat CY metric is not known in analytic form, its associated volume form  can be constructed explicitly (up to a constant) from the expression for $\Omega$. 

\subsection{Transition functions}\label{sec:transition}

The affine coordinates on each patch of the (toric) ambient space patch can be written as functions of the homogeneous coordinates $x^i$ (see Appendix~\ref{app:AffinePatches}). More explicitly, for two ambient space patches $\mathcal{U}$ and $\mathcal{V}$ with affine coordinates $u^\mu$ and $v^\nu$,\footnote{We caution the reader not to confuse these affine coordinates $u^\mu$ and $v^\nu$ with e.g.~the toric vertices discussed in other sections of this paper.} respectively, we can write $u^\mu=u^\mu(x)$ and $v^\nu=v^\nu(x)$, where the functional dependence on $x^i$ is determined by Eq.~\eqref{eq:affinecoordinates}. The transition function $v^\nu=v^\nu(u)$ can then be determined by matching the functional dependence on the homogeneous coordinates $x^i$. The transition matrix is then simply the Jacobian of this coordinate change,
\begin{align}
(T_{\mathcal{U}\mathcal{V}})^\nu_\mu=\frac{\partial v^\nu}{\partial u^\mu}\,.
\end{align}

\subsection{Constructing K\"ahler forms}\label{sec:kmetric}
A further ingredient which we require explicitly is a basis $J_{\alpha}$, where $\alpha=1,2,\ldots, h^{1,1}(X)$, of the second cohomology. This basis is used to expand the K\"ahler form $J_\FS$, as in Eq.~\eqref{JFSdef}, and the K\"ahler parameters $t^\alpha$ are defined relative to it. For simplicity, we focus on simplicial K\"ahler cones so that the $J_{\alpha}$ can be chosen as generators of the K\"ahler cone. Since these forms can be written in terms of K\"ahler potentials 
\begin{equation}\label{Jalphadef}
 J_{\alpha}=\frac{i}{2\pi}\partial\bar{\partial}\ln(\kp_{\alpha})
\end{equation}
our task is to construct the $\kp_{\alpha}$. 

Recall that for our purpose of machine learning Ricci-flat metrics on Calabi-Yau manifold, we need to generate a random sample of points on the CY. We will discuss the method we employ for this task in section~\ref{sec:torsamp}. A crucial point of this method is that it relies on properties of Fubini-Study K\"ahler metrics on projective spaces. Our objective in this section will therefore be to construct K\"ahler potentials $\kp_\alpha$ that are 'inherited' from ambient space FS potentials on $\bP^{r_\alpha}$, both for CICY and KS CY manifolds. For CICYs, this is a standard procedure. For KS CYs, the construction is less well known, so we will explain it in some detail.

As already mentioned above (and as explained in detail in Appendix~\ref{app:ToricGeometryIntro}), toric ambient spaces have homogeneous coordinates $x_i$. The vanishing locus of each $x_i$ defines a toric divisor, $D_i=\{x_i=0\}$, and the K\"ahler cone generators $J_\alpha$ may be expanded in terms of these divisors: $J_\alpha=\sum_i c_i^\alpha D_i$.  Denoting the space of sections of the (nef) line bundle dual to the K\"ahler cone divisors by $H^0(J_{\alpha})$, we can find a monomial basis $\{s_j^{(\alpha)}\}$, $j=0,1,\ldots, h^0(J_{\alpha})-1:=r_\alpha$, of this space following the methods outlined in appendix~\ref{app:Sections}. In particular, we have 
\begin{equation} \label{eq:tormap}
s_j^{(\alpha)} = \prod_{i=1}^n x_i^{\langle v_i,w_j\rangle+c_i^\alpha} 
\end{equation}
which expresses the monomial basis in terms of the toric coordinates.  Here $v_i$ are the vectors (1d cones) in the fan associated to the toric variety, and subsets of these  span the top-dimensional cones of the lattice. The $w_j$ are vectors in the dual top-dimensional cones of the  dual lattice (see Appendix~\ref{app:ToricGeometryIntro} for the precise definition). In general, the dimension of the space of sections will be larger than the number of homogeneous coordinates $x_i$, implying that~\eqref{eq:tormap} is a non-invertible relation.

By the theorem quoted in Section~\ref{sec:LineBundlesSections} below~\eqref{eq:divisorsections}, nef divisors are basepoint free, which means that the sections of the nef line bundles cannot all vanish at any given point in the CY. This means that the section monomials $s_j^{(\alpha)}$ can serve as homogeneous coordinates of a projective space $\bP^{r_\alpha}$. The relations given in eq.~\ref{eq:tormap} thus correspond to maps $\Phi_\alpha$ mapping the toric coordinates onto the coordinates of these projective spaces
\begin{align}\label{Phidef}
\Phi_\alpha:~~~[x_0: x_1:\ldots] \quad\to\quad [s_0^{(\alpha)}:s_1^{(\alpha)}:\ldots:s_{r_\alpha}^{(\alpha)}]\,.
\end{align}
 Note that in the case where the ambient space is just a projective space $\bP^{n}$ with homogeneous coordinates $[x_0:x_1:\ldots:x_{n}]$, the sections of the K\"ahler cone generator are simply the monomials $s_i=x_i$, and the morphism $\Phi$ would be the identity map $\Phi:\bP^{n}\to\bP^{n}$. In the toric case, the map is non-trivial, and there are typically more sections than toric coordinates, leading to relations among the sections. 

We can express the K\"ahler potentials $\kp_{\alpha}$ via inner products of sections,
\begin{align}\label{eq:KahlerRho}
\kp_{\alpha}= s^{(\alpha)\,\dagger}\cdot \mathbbm{1}\cdot s^{(\alpha)}\,,
\end{align}
where $s^{(\alpha)}$ is a vector containing the section monomials $s_i^{(\alpha)}$ and $\mathbbm{1}$ is the identity matrix (or, as a straightforward generalization, any hermitian, non-degenerate matrix) and the inner product induces a notion of orthogonality (with respect to this matrix) as will become important later when sampling points. Since not all $s_\alpha$ can vanish simultaneously, ${\cal K}$ is positive and is, hence, a sensible argument of the  logarithm in Eq.~\eqref{Jalphadef}. In the case of a projective ambient space $\bP^n$, this K\"ahler potential is the standard FS K\"ahler potential
\begin{align}\label{eq:FSMetricCase}
\kp_\FS=\sum_{i=0}^n |s_i|^2 = \sum_{i=0}^n |x_i|^2\,.
\end{align}

More generally, inserting the monomial basis~\eqref{eq:tormap} in~\eqref{eq:KahlerRho}, one obtains a K\"ahler potential for each K\"ahler cone generator of a toric variety
\begin{align}
\label{eq:rhoToric}
\kp_{\alpha}=\sum_j|s_j^{(\alpha)}|^2=\sum_{\stackrel{w_j\in M}{\langle v_i,w_j\rangle+c_i^\alpha\geq0~\forall i}}\left|\prod_ix_i^{\langle v_i,w_j\rangle+c_i^\alpha}\right|^2\,,
\end{align}
expressed in terms of homogeneous coordinates. Clearly, this is the standard Fubini-Study metric on the projectivized section space $\mathbb{P}H^0(J_{\alpha})\cong \bP^{r_\alpha}$, pulled back to the toric variety using the map~\eqref{Phidef}. 

The above results for $\kp_\alpha$ are written in terms of the toric coordinates $x_i$. However, for subsequent applications we would like to re-write these in terms of affine coordinates on a patch. This can be accomplished by dividing $\kp_\alpha$ in~\eqref{eq:rhoToric} by a monomial that is nowhere-vanishing in a chosen patch and that has the same toric scaling as $\kp_\alpha$. It turns out, there is a unique choice $s_*^{(\alpha)}$ for this non-vanishing monomial, specified by a vector $w_*$ in the dual lattice. Let us denote the affine patches by $\mathcal{U}_r$, with associated dual cones $\sigma_r^*$ and affine coordinates $z_\mu$. Then, as discussed in Appendix~\ref{app:AffinePatches}, the toric and affine coordinates are related by
\begin{align}
z_\mu=\prod_{i=1}^n x_i^{\langle v_i,w_\mu \rangle}\,.
\end{align}
We introduce the index set $\mathcal{I}_r$ of indices whose associated coordinates generate the corresponding cone $\sigma_r$. Then, the affine version of $\kp_\alpha$ in~\eqref{eq:rhoToric} in a  patch $\mathcal{U}_r$ can be written as
\begin{align}
\label{eq:kappapatch}
\kp_{\alpha,\;r}=1+\sum_{\stackrel{w_j\in \sigma_r^*}{\langle v_i,w_j+w_*\rangle+c_i^\alpha\geq0~\forall i\not\in\mathcal{I}_r}}\left|\prod_{\mu\in\mathcal{I}_r} z_\mu^{\langle v_\mu,w_j\rangle}\right|^2\,,
\end{align}
where the dependence on the patch $r$ has been made explicit. For the case where the ambient space is simply a projective space, this reduces to the standard (affine version of the) Fubini-Study K\"ahler potential, $\kp_\FS=1+\sum_\mu |z_\mu|^2$.

%%%%%%%%%%%%%%%%%%%%%%%%%%%%%%%%%%%%%%%%%%%%%%%%%%%%%%%%%%%%%%%%%%%%%%%%%%%%
\subsection{Numeric integration}
Numeric integration can be performed using Monte-Carlo methods (for a method developed recently based on rejection sampling from tropical densities see~\cite{Borinsky:2022aaa}). The basic goal is to be able to numerically evaluate integrals of the form
\begin{align}\label{intf0}
 \int_X\text{d}\vCY\; f\,.
\end{align}
for a function $f$ on a CY manifold $X$ (of CICY or KS type). To do this, we require a sample of points $p_i\in X$, where $i=1,\ldots ,N$, which are distributed according to a known measure, denoted by $dA$, which we discuss in detail below. In terms of the so-called weights (or masses), customarily defined as
\begin{align}
 w_i=\left.\frac{\text{d}\vO}{dA}\right|_{p_i}\; .
\end{align}
and the function values $f(p_i)$, the integral~\eqref{intf0} can be approximated by
\begin{align}\label{intf1}
  \int_X\text{d}\vCY\; f=\int_X dA\,\frac{\text{d}\vCY}{dA}\; f\simeq\frac{\kappa}{6N}\sum_{i=1}^N w_i f(p_i)\; ,
\end{align}
where~\eqref{dVCYdVO} has been used. On the other hand, since the Ricci-flat CY metric is calculated numerically, there is no real need to rely on the measure $\text{d}\vO$ and the integral~\eqref{intf0} can alternatively be approximated by
\begin{align}\label{intf2}
\int_X\text{d}\vCY\; f\simeq \frac{1}{N}\sum_{i=1}^N\tilde{w}_i\, {\rm det}(g_\CY(p_i))\, f(p_i)\;,\qquad
\tilde{w}_i=\left.\frac{d^6y}{dA}\right|_{p_i}\; .
\end{align}
In the following we refer to the quantities $\tilde{w}_i$ as the auxiliary weights. The approximation~\eqref{intf2} has a certain advantage in that it does not contain any unknown coefficient, such as $\kappa$ in Eq.~\eqref{intf1}, which need to be fixed by a reference calculation.

\subsection{Sampling points on toric varieties} \label{sec:torsamp}
We now discuss a method for generating the sample points $p_i$ with a known distribution $dA$ on the CY $X$. In principle, this could be done by a Markov chain Monte-Carlo method using, for example, the Fubini-Study measure for $dA$, so $dA=\text{d}\vFS$. Here, we pursue a different approach based on a theorem by Shiffman and Zelditch~\cite{Shiffman:1999aaa}, further generalized and explained in~\cite{Douglas:2006rr,Braun:2007sn}. This provides us with a viable method for ambient spaces which are products of projective spaces, and hence applies straightforwardly to CICYs. Importantly, using the techniques introduced in section~\ref{sec:kmetric}, which allow us to write the toric K\"ahler potential in terms of a FS potential in the projectivized section space $\mathbb{P}H^0(J_\alpha)\cong\bP^{r_\alpha}$ of the K\"ahler cone generators~$J_\alpha$, we can apply the same method to the toric case, as we will now explain.

To construct $dA$, we first compute the sections~\eqref{eq:tormap} and construct the associated FS K\"ahler potentials $\kp_\alpha$ as in Eq.~\eqref{eq:KahlerRho}. Shiffman and Zelditch now state that the zeros of random sections are distributed w.r.t.\ to the (known) FS measure constructed from~$\log(\kp_\FS)=\log\sum_{i=0}^n |s_i|^2$. Hence, we need to find zeros of random sections that are also lying on the CY and express these in terms of the homogeneous toric coordinates $x_i$.

First, we need to clarify what random section actually means in this context. As mentioned in section~\ref{sec:kmetric}, the scalar product induces a notion of orthogonality (and hence independence) of sections. Since we chose the standard scalar product, the monomial sections are mutually orthogonal, and a random section $S_{\alpha}$ of the K\"ahler cone generator $J_\alpha$ is constructed from this orthogonal basis $s_{j}^{\alpha}$ with i.i.d.~Gaussian coefficients $a_j^{(\alpha)}\sim\mathcal{N}(0,1)$
\begin{align}\label{eq:RandomSection}
S_{\alpha}=\sum_{j=0}^{r_\alpha} a_j^{(\alpha)} s_{j}^{\alpha}\; .
\end{align}
Since we want points on a CY $n$-fold, we need to construct $n$ random sections $S_{\alpha,p}$, $p=1,\ldots,n$, and intersect their zero locus with the CY hypersurface, specified by its defining equation. We are relatively free in choosing the K\"ahler cone generators (that is, the indices~$\alpha$ for each $p=1,\ldots,n$) for which we construct the random sections, as long as we are imposing not more than $r_\alpha$ equations for any given $\alpha$.

We can use this to construct a point sample with known distribution and measure $dA$. Since we want to work with the toric coordinates $x_i$ instead of the sections $s_j^{\alpha}$, we need to express the sections in terms of the $x_i$. As mentioned around~\eqref{Phidef}, the problem is that the maps $\Phi_\alpha$ from $x_i$ to $s_j^\alpha$ are not invertible. Indeed, there are non-linear relations (higher syzygies) among the $s_j^\alpha$. Mathematically, this means that we have to construct a polynomial ring in $x_i$ and $s_j^\alpha$ and add the ideals that describe the maps $\Phi_\alpha$, the hypersurface equations for the CY, and $n$ random sections $S_{\alpha,p}$. Then, one finds the elimination ideal in the quotient ring modulo the relations among the sections to express the $s_j^\alpha$ in terms of the $x_i$. We carry this out in a two-step procedure. We solve the $n$ equations $S_{\alpha,p}=0$ subject to the relations among the sections (we will discuss how to find these relations in the next paragraph). We then express the remaining $s_j^\alpha$ in terms of the $x_i$ as in~\eqref{eq:tormap} and impose the hypersurface equations (which are equations in the $x_i$). Solving this system of equations then gives a collection of homogeneous coordinates of points on the CY for each choice of $n$ random sections. Moreover, these points are distributed with respect to the auxiliary measure $dA$. For a CY three-fold,
\begin{align}
dA=\Phi_\alpha^*(J_{\alpha})\wedge\Phi_\beta^*(J_{\beta})\wedge \Phi_\gamma^*(J_{\gamma})\,,
\end{align}
where the indices $\alpha$, $\beta$, $\gamma$ correspond to the indices of the random sections $S_{\alpha,p}$, $p=1,2,3$, that have been chosen above. Repeating this process for many random sections gives us the point sample we are after. While we appreciate that this is somewhat involved, it is absolutely crucial for our purposes. Any meaningful numerical integration relies on being able to generate points efficiently and distributed according to a known measure, $dA$. This is precisely what the above method allows us to do.

Finally, we need to explain how to derive the relations among the sections. One way to do this is to perform a primary decomposition of the defining ideal of the aforementioned quotient ring and use the elimination ideal to express the ideal generators, as obtained from the primary decomposition, in terms of the sections. This method is general, and we implemented it in Sage (via Singular). However, it requires computing Groebner bases, which can be computationally costly. 

Alternatively, we can do the following: We are interested in relations of the form\footnote{In principle, we are free to use any coefficients, but a different choice would modify the parameterization of the complex structure as coefficients of the section monomials of the anti-canonical bundle that appear in the hypersurface that defines the CY.}
\begin{align}\label{eq:SectionRelationsRaw}
\prod_{I}s_{I}^{f_I} = \prod_{J}s_{J}^{g_J}\quad\Leftrightarrow\quad \prod_{I}s_{I}^{h_I}=1\,,
\end{align}
where we have introduced the multi-index $I=(i,\alpha)$, and likewise for $J$, and $f_I,g_J$ are non-negative integers. At a generic point in section space, we can divide by the right-hand side and write the equation as done in the final step, where the $h_I$ are integers (positive or negative) which encode the relations between the sections. Now recall from~\eqref{eq:divisorsections} or~\eqref{eq:tormap} that the sections are given in terms of the toric coordinates as
\begin{align}\label{eq:relSections}
s_J=\prod_a x_a^{\langle v_a,w_j^{(\alpha)}\rangle+c_{i}^{(\alpha)}}=:\prod_a x_a^{M_{a,J}}\,,
\end{align}
where the last step defines the matrix of exponents $M_{a,J}$. Inserting this into~\eqref{eq:SectionRelationsRaw}, we get
\begin{align}
\sum_{I} M_{a,I} h_I = 0\; .
\end{align}
This means the vectors $h_\alpha$ can be found as the (primitive integer) basis of the kernel of $M_{a,I}$. This approach based on lattice methods avoids potentially time-consuming Groebner basis calculations and is the default method used by the code.

Finally, we want to point out the following: At first, one might think that the system of equations comprising the CY hypersurface equation plus the equations that determine the sections in terms of the toric coordinates is simply a CICY in a product of projective ambient spaces $\otimes_{\alpha=1}^{h^{1,1}(X)}\mathbbm{P}^{r_\alpha}$. However, if we naively count the dimensions of the resulting variety, we would find that it is in general smaller than $n$. Moreover, the degrees of the equations do not add up to the toric scalings plus one, so the hypersurface also seems to not be the anti-canonical one. However, the point is that the CY is \emph{not} a complete intersection, precisely because of the relations of the sections~\eqref{eq:relSections}. So another way of summarizing the procedure is that we have found an expression for the original CY $n$-fold $X$ as a non-complete intersection CY in a product of projective ambient spaces $\otimes_{\alpha=1}^{h^{1,1}(X)}\mathbbm{P}^{r_\alpha}$, which allows us to generate point samples with known distribution on $X$.

%%%%%%%%%%%%%%%%%%%%%%%%%%%%%%%%%%%%%%%%%%%%%%%%%%%%%%%%%%%%%%%%%%%%%%%%%%%%
\section{Hermitian Yang-Mills equations}\label{sec:hym}
Line bundles $L\rightarrow X$ over CY manifolds $X$ and their Hermitian Yang-Mills (HYM) connections provide an interesting application of numerical CY metrics. A vanishing slope is a necessary (and sufficient) condition for the existence of a HYM connection on a line bundle. Hence, meaningful numerical calculation of the HYM connection rely on being able to determine the Ricci-flat metric at prescribed points in K\"ahler moduli space - precisely one of the capabilities we are developing in the present work. In this section, we briefly review the underlying mathematics and set up the relevant formalism for our purposes. Numerical results for HYM connections will be presented in Section~\ref{sec:results}.

\subsection{Mathematical background}\label{sec:hymback}
Line bundles ${\cal O}_X(k)$ over CY three-folds $X$ and their slope have already been introduced in Section~\ref{sec:linebundles}. If $H$ is a hermitian bundle metric on ${\cal O}_X(k)$ we have an associated (Chern) connection with gauge field and curvature given by
\begin{equation}
 A=\partial\ln (H)\;,\qquad F=\bar{\partial}A=\bar{\partial}\partial\ln(H)\; .
\end{equation}
The above connection is said to be a HYM connection if its field strength satisfies the HYM equation
\begin{equation}\label{HYMeq}
 g_\CY^{a\bar{b}}F_{a\bar{b}}=0\quad\Leftrightarrow\quad J_\CY\wedge J_\CY\wedge F=0\; ,
\end{equation}
where $g_\CY$ is the (Ricci-flat) CY metric and $J_\CY$ the associated K\"ahler form. A necessary and sufficient condition for the existence for a HYM connection on ${\cal O}_X(k)$ is that its slope, as defined in Eq.~\eqref{eq:slope}, vanishes, that is, $\mu_t({\cal O}_X(k))=0$. Checking this condition numerically is, therefore, the first step in a numerical calculation of the HYM connection.

\subsection{Computing HYM connections}\label{sec:hymconnection}
Provided the slope $\mu_t({\cal O}_X(k))$ vanishes so that a HYM connection exists, how can this connection be computed explicitly? The first step is to start with a ``Fubini-Study'' reference bundle metric on ${\cal O}_X(k)$ defined by
\begin{equation}\label{eq:FFS}
 H_\FS=\prod_\alpha \kp_\alpha^{-k_\alpha}\;,\qquad F_\FS=\bar{\partial}\partial\ln(H_\FS)\; ,
\end{equation}
with the quantities $\kp_\alpha$ constructed in Section~\ref{sec:kmetric}. This reference metric can be modified to 
\begin{equation}\label{eq:FCY}
 H=e^\beta H_\FS\;,\qquad F=\bar{\partial}\partial\ln(H)=F_\FS+\bar{\partial}\partial\beta
\end{equation} 
where $\beta$ is a function on $X$ to be determined.  Noting that the scalar Laplacian $\Delta_\CY$ associated to the Ricci-flat metric $g_\CY$ on $X$ can be written as $\Delta_\CY=2g^{a\bar{b}}_\CY\partial_a\bar{\partial}_{\bar{b}}$, the HYM equation~\eqref{HYMeq} for the field strength $F$ from Eq.~\eqref{eq:FCY} turns into
\begin{equation}\label{betaeq}
 \Delta_\CY\beta=\rho_\CY
\end{equation}
where the source $\rho_\CY$ has been defined in Eq.~\eqref{eq:rhodef}.  We note that the integrability condition for this equation amounts to $\rho_\CY$ integrating to zero over the CY manifold which, from Eq.~\eqref{eq:slope}, is equivalent to the slope vanishing. Provided this is the case, Eq.~\eqref{betaeq} has a solution $\beta$ (unique up to an additive constant) which, inserted into Eq.~\eqref{eq:FCY}, leads to the Hermitian bundle metric underlying the HYM connection.

In order to solve Eq.~\eqref{betaeq} numerically we introduce the usual scalar product
\begin{equation}
 \langle f|h\rangle:=\int_X\bar{f}\wedge\star_g h=\int_Xd^6y\sqrt{g}\,\bar{f}\,h
\end{equation}
on the space $L^2(X)$ and consider a (not necessarily ortho-normal) basis of functions denoted by $f_I$ or simply by $|I\rangle$. If we define the scalar products
\begin{equation}\label{matel}
G_{IJ}=\langle I|J\rangle\;,\quad \Delta_{IJ}=\langle I|\Delta_\CY|J\rangle\;,\quad \tilde{\beta}_I=\langle I|\beta\rangle
 \;,\quad \rho_I=\langle I|\rho\rangle\; ,
\end{equation} 
it follows that
\begin{equation}
 |\beta\rangle=\sum_I\beta_I |I\rangle\quad\mbox{where}\quad \beta_I=\sum_J(G^{-1})_{IJ}\,\tilde{\beta}_J
\end{equation}
and the Laplace equation~\eqref{betaeq} can be re-written as
\begin{equation}\label{betaeqla}
 \Delta{\boldsymbol\beta}={\boldsymbol\rho}\; .
\end{equation}
Here, $\Delta$ is the matrix with entries $\Delta_{IJ}$ and we have introduced the vectors ${\boldsymbol\beta}=(\beta_I)$ and ${\boldsymbol\rho}=(\rho_I)$. The kernel of $\Delta$ is non-trivial and consists of the vectors ${\boldsymbol\beta}$ which correspond to constant functions on $X$. This means $\Delta$ is not surjective and for a solution to exist it is necessary that ${\boldsymbol\rho}\in{\rm Im}(\Delta)$. This condition is of course nothing else but the integrability condition for the Laplace equation~\eqref{betaeq} which amounts to the vanishing of the slope. So as long as the slope vanishes, Eq.~\eqref{betaeqla} has a solution.

In practice, we have to work with a suitably chosen finite set of functions $f_I$ and things are not quite so straightforward. For a  finite function set we should expect errors and we cannot expect that ${\boldsymbol\rho}\in{\rm Im}(\Delta)$. This means the finite-dimensional version of Eq.~\eqref{betaeqla} typically does not have a solution and a naive attempt of solving the equation with standard methods of linear algebra fails. To remedy this problem we decompose ${\boldsymbol\rho}$ as ${\boldsymbol\rho}={\boldsymbol\rho}^\parallel+{\boldsymbol\rho}^\perp$ into the projection ${\boldsymbol\rho}^\parallel$ onto ${\rm Im}(\Delta)$ and the orthogonal complement ${\boldsymbol\rho}^\perp$. Another way of stating the problem is that ${\boldsymbol\rho}^\perp$ is typically non-zero in the finite-dimensional approximation. However, provided the slope vanishes and the finite-dimensional approximation is sufficiently accurate, we expect that $|{\boldsymbol\rho}^\perp|/|{\boldsymbol\rho}^\parallel|\ll 1$.  Assuming this is indeed satisfied we can safely replace Eq.~\eqref{betaeqla} by
\begin{equation}\label{betaeqla2}
 \Delta{\boldsymbol\beta}={\boldsymbol\rho}^\parallel\; ,
\end{equation}
which does have a solution.

In summary, our procedure involves computing the quantities $\Delta_{IJ}$ and $\rho_I$  in Eq.~\eqref{matel}  by numerical integration over $X$ for a finite set of function $f_I$ and then solving the linear algebra problem~\eqref{betaeqla2}.
Given a solution ${\boldsymbol\beta}=(\beta_I)$, the function $\beta$ is then approximated by $\beta=\sum_I\beta_I f_I$. 

A suitable set of function $f_I$ can be constructed by starting with a positive line bundle ${\cal O}_X(l)$. If we denote by $s_p$, where $p=1,\ldots ,h^0({\cal O}_X(l))$, a basis of its sections we can define the functions
\begin{equation}\label{eq:fbasis}
 f_{pq}=\frac{s_p\bar{s}_q}{\prod_\alpha {\mathcal S}_\alpha^{l_\alpha}}\; ,
\end{equation}
where ${\mathcal S}_\alpha$ is the homogeneous version of $\kp_\alpha$. Of course, the higher the degree $l$, the larger the space spanned by the $f_{pq}$ and the more accurate we expect the approximation to be.  As a measure of the accuracy with which the HYM equation~\eqref{betaeq} is solved we define
\begin{equation}\label{eq:HYMacc}
\sigma_H = \frac{\int_Xd^6y\sqrt{g_\CY}\,|\Delta_\CY\beta-\rho_\CY|}{\int_X d^6y\,\sqrt{g_\CY}\, |\rho_\CY|}\; .
\end{equation}
which we will refer to as the 'HYM measure' in the future.

%%%%%%%%%%%%%%%%%%%%%%%%%%%%%%%%%%%%%%%%%%%%%%%%%%%%%%%%%%%%%%%%%%%%%%%%%%%%
\section{Computational background}\label{sec:compbackground}

The open source library \texttt{cymetric} consists of point generators for CICY and KS Calabi-Yau manifolds and five metric models that learn the Ricci-flat CY metrics using the generated points as input. In this section, we describe these networks and their training in more detail. After a brief general introduction to neural networks, we describe how the networks of \texttt{cymetric} are constructed and how their learning is controlled by custom loss functions that encode the mathematical constraints CY metrics must satisfy.

\subsection{Introduction to neural networks}\label{sec:nns}

This section provides a quick review of the building blocks of neural networks. The literature on machine learning and neural networks is vast, and includes many good introductory texts. For the reader with a background in (theoretical) high-energy physics,  Ref.~\cite{Mehta:2018dln} provides a very good introduction to the topic, and Ref.~\cite{Ruehle:2020jrk} gives a topical discussion on the applications of modern data science techniques to string theory. 

A neural network (NN) is a collection of nodes, which are organized in several layers. Two of these layers, the input and output layers, are visible to the user, the others are hidden. NNs with many nodes in each layer are referred to as wide, and NNs with many (more than one) hidden layers are called deep.  Given an input vector $v_0\in\mathbb{R}^{n_0}$, the NN processes the data  by transmitting it from one layer to the next, in a sequence of mappings $v_0\mapsto v_1\mapsto v_2\mapsto\cdots\mapsto v_N$, where $v_k\in\mathbb{R}^{n_k}$. 
In more detail, for a fully-connected NN, the output $v_{k}$ of layer $k$  is related to its input $v_{k-1}$ by 
\begin{equation}\label{eq:NNlayer}
v_{k}=\sigma_k(W_{k} v_{k-1} + b_{k}) \; ,
\end{equation}
where the $n_{k}\times n_{k-1}$ matrix $W_{k}$ contains the {\it weights} and the vector $b_{k}\in\mathbb{R}^{n_{k}}$ the {\it biases} of the $k^{\rm th}$ layer. The functions $\sigma_k$ acts on each component of their vectorial arguments and are called the \emph{activation functions}. Altogether, the neural network represents a family $f_\theta:\mathbb{R}^{n_0}\rightarrow\mathbb{R}^{n_N}$ of functions which is given by iterating Eq.~\eqref{eq:NNlayer}. Here, $\theta$ collectively denotes the weights and biases in all layers. 
 
Training of the neural network is based on a set of input vectors $\{v_0\}$. The first step of any machine learning experiment is to randomly split this data into a training and a validation set (typical training-to-validation ratios are around 9:1, or validation split 0.1). The training set is used to train the NN by adjusting the weights and biases $\theta$, a process that is guided either by certain loss functions or a reward system provided by the user (see below for the loss functions relevant to our setting). The training data may also be further subdivided into (mini-) batches to improve training. The division of data into batches reduces the risk of the NN getting trapped in  local minima of the loss function, thus increasing the chance of finding solutions that generalize well. This is also sometimes necessary, especially for networks that perform memory-intensive operations (such as differentiation with respect to the input data in our case) in order to compute loss functions. Such networks operate in epochs, where each epoch constitutes one run over all batches.
 
In machine learning, one typically distinguishes different setups such as supervised, unsupervised, or reinforcement learning. We are dealing with a version of supervised machine learning. While in classical supervised ML, the input \textit{and} output is known for the training set, we only have inputs (the points on the CY at which to compute the metric), together with a set of constraints that the neural network output (that is, the metric) has to satisfy. Hence, the task is more akin to physics informed learning or optimization than classical 'supervision' with given input-output pairs. We check how well the output meets these constraints by constructing error measures that encode them. Importantly, given the split of the initial data, the quality of the output can be assessed on both the training and validation sets. For the validation set, this check is performed by applying the trained network to the validation data.

From this brief summary, it is clear that designing a NN to perform a certain task involves choosing a number of {\it hyperparameters}, that specify the network's properties. For fully connected NNs, this amounts to specifying the width and depth of the network, which activation functions to use,  and the number of batches and epochs. More advanced network components, such as convolutional or dropout layers (see, for example, Ref.~\cite{Ruehle:2020jrk} for a recent discussion), require additional hyperparameters. For our task, previous studies~\cite{Anderson:2020hux,Jejjala:2020wcc,Larfors:2021pbb} have shown that a simple fully connected NN exhibits good performance. Consequently, we  focus on this setting, leaving studies involving more advanced NN architectures for the future.  We build the networks using TensorFlow~\cite{tensorflow2015-whitepaper}, an ML library with functionalities that suit our needs. In particular, TensorFlow allows to construct the NNs in a sequential manner, and has differentiation functions that allow us to compute derivatives with respect to the input data.

\subsection{Network architectures}\label{sec:arch}

The basic idea underlying machine learning of CY metrics is to use a NN whose associated functions $f_\theta$ represent metrics on the manifold. In other words, the NN input consists of a point $p\in X$ on the CY manifold and its output represents a metric $g(p)$ at this point. There are a number of concrete realizations of this idea. Since CY three-folds $X$ are complex manifolds, their metrics $g(p)$ at each point $p\in X$ can be written, relative to a local choice of complex coordinates, as a Hermitian $3\times 3$ matrix. The first, and most  obvious, approach is then to let the NN predict the nine independent entries (three real entries on the diagonal, and three complex entries on the off-diagonal) of this matrix. While this is possible, it does not take advantage of the mathematical knowledge we have about CY manifolds. For example, equation~\eqref{eq:MA} shows that the CY metric  is given by an exact correction to some reference K\"ahler metric $\gFSX$. Moreover, by constructing CY manifolds as a hypersurfaces or complete intersection in an ambient space, one can construct the metric $\gFSX$ explicitly by pullback from the ambient space $\mathcal{A}$.

The  \texttt{cymetric}  package realizes five choices for how the metric $\gpred$ predicted by the NN is related to the function $g_{\text{NN}}$ that the NN actually represents. These possibilities are summarized in Table~\ref{tab:ansatz}.
\begin{table}[t]
\centering
\begin{tabular}{@{}|c|c|@{}}
	\hline
	Name& Ansatz \\ \hline \hline
	Free & $\gpred = g_{\text{NN}}$ \\
	Additive & $\gpred = g_{\text{FS}} + g_{\text{NN}}$ \\
	Multiplicative, element-wise & $\gpred = g_{\text{FS}} + g_{\text{FS}} \odot g_{\text{NN}}$ \\
	Multiplicative, matrix & $\gpred = g_{\text{FS}} + g_{\text{FS}} \cdot g_{\text{NN}}$ \\
	$\phi$-model & $\gpred = g_{\text{FS}} + \partial \bar{\partial} \phi$ \\
	\hline
\end{tabular}
\caption{Different Ans\"atze for the neural network prediction of the Ricci-flat metric.}
\label{tab:ansatz}
\end{table}
The first and most obvious choice, $\gpred=g_{\text{NN}}$, has been included for reference but is by no means the optimal one. A metric is required to be non-singular and this condition can easily be violated for a randomly initialized or stochastically trained NN. A NN which 'accidentally' represents a singular or near-singular metric can lead to numerical problems. Also, for this choice, the NN has to cope with the entire numerical variation of the metric $\gpred$. Both problems can be solved, or at least alleviated, by writing $\gpred=\gFSX+\text{correction}$, and the four other possibilities in Table~\ref{tab:ansatz} are of this form. Indeed, using the non-singular metric $\gFSX$ as a background makes accidentally generating singular metrics less likely (and, if there is an actual singularity in the space, the reference \FS metric will already have this feature). Furthermore, under the plausible assumption that the entries of $\gCY-\gFSX$ are typically smaller than the ones in $\gCY$, a NN dealing with a metric correction involves smaller numbers, likely leading to enhanced numerical stability. As Table~\ref{tab:ansatz} shows, $\gpred$  may be defined additively, $\gpred = \gFSX + g_{\text{NN}}$, via element-wise multiplication, $\gpred =\gFSX + \gFSX \odot g_{\text{NN}}$, or via matrix multiplication, $\gpred = \gFSX + \gFSX \cdot g_{\text{NN}}$. Finally,  using the full information contained in Eq.~\eqref{eq:MA}, we may set $\gpred = \gFSX + \partial \bar{\partial} \phi$, where the NN represents the real function $\phi$.

While we have already argued that the free NN is likely inefficient, there is no  telling which of the other four Ans\"atze in Table~\ref{tab:ansatz} will be most efficient in learning the Ricci-flat metric on a given CY manifold. For example, while the $\phi$-model automatically gives a K\"ahler metric, it requires two additional derivatives on the input, which come at a computational cost. For this reason, the  \texttt{cymetric} package makes all these options available for the user to explore. In the ensuing section, we will compare the performance of the different NNs on the quintic. 

In each experiment the NN is trained on input data which consists of a set $\{p_i\}$ of points on the CY manifold. These points are generated using the method described in section \ref{sec:torsamp}. They are, by the theorem of Shiffman and Zelditch, distributed with respect to a measure $d A$, which is inherited from the ambient space. This allows to compute any integrals used in the training and validation processes as weighted sums, following Eqs.~\eqref{intf1} and \eqref{intf2}.

As discussed above, the learning of NNs is governed by the minimization of loss functions. A CY metric must satisfy a number of mathematical constraints, listed in section \ref{sec:background}. We encode these restrictions in a custom loss function
	\begin{align}
		\label{eq:loss}
		\cL &= \alpha_1 \cL_{\text{MA}} + \alpha_2 \cL_{\text{dJ}} + \alpha_3 \cL_{\text{transition}} + \alpha_4 \cL_{\text{Ricci}} + \alpha_5 \cL_{\text{Kclass}}
	\end{align}
where $\alpha_i$ are hyperparameters of the NN with default value $\alpha_i = 1.0$, and the individual loss contributions are defined as follows:
\begin{itemize}
\item {\it Monge-Amp\`ere loss} $\cL_{\text{MA}}$\\
The Monge-Amp\`ere loss is defined as
\begin{equation} \label{eq:mariloss}
\cL_{\text{MA}} = \left|\left| 1 - \frac{1}{\kappa} \frac{\det \gpred}{\Omega \wedge \bar\Omega}\right|\right|_n  \;  ,
\end{equation}
and it ensures that  the metric satisfies the Monge-Amp\`ere equation~\eqref{eq:MA}, as required by the Calabi--Yau theorem.  This loss may be computed with any $L_n$ norm, as indicated by the subscript $n$ (default is $n = 1$).\footnote{The choice of $L_n$ norm will affect the training of the network. A high $n$ pushes the network to reduce outliers, that is, large loss function contributions that are localized at a few points. With a low $n$, the NN will instead strive to reduce the loss function of all points with equal weight. From our experiments, the default values chosen in \texttt{cymetric} lead to good performance on the manifolds we have studied.}
\item {\it Ricci loss} $\cL_{\text{Ricci}}$\\
The Ricci loss is defined by
\begin{equation}
\cL_{\text{Ricci}} = ||R||_n = \left|\left|\partial \bar\partial \ln{\det\gpred}\right|\right|_n \;  ,
\end{equation}
and it ensures that  the metric has vanishing Ricci scalar.\footnote{Recall that for K\"ahler metrics, the Ricci scalar simplifies to
$R =  g^{i\bar{j}}\partial_{i} \bar{\partial}_{\bar{j}} \log \det (g)$.} This is a necessary condition for the metric to satisfy the stronger CY requirement of having vanishing Ricci tensor.  Again, this loss function may be computed with different $L_n$ norms, with default $n = 1$.
\item {\it K\"ahler loss} $\cL_{\text{dJ}}$\\
The K\"ahler loss, defined by
\begin{equation} \label{eq:kloss}
\cL_{\text{dJ}} = \sum_{ijk}  \left|\left|\Re{c_{ijk}}\right|\right|_n +  \left|\left|\Im{c_{ijk}}\right|\right|_n \;  ,
\end{equation}
where $c_{ijk} = g_{i\bar{j},k} - g_{k\bar{j},i}$ and $g_{i\bar{j},k} = \partial_k g_{i\bar{j}}$ ensures that $d J =0$, so that the metric is K\"ahler. The default for the $L_n$ norm is $n = 2$. 
\item {\it Transition loss} $\cL_{\text{transition}}$\\
The {\it transition loss function} ensures that  $g_{i\bar{j}}$ transforms as a complex tensor under coordinate transformations. For the additive and multiplicative networks, this is implemented as
\begin{equation}
\cL_{\text{transition}} = \frac{1}{d} \sum_{\mathcal{U},\mathcal{V}} \left|\left|\gpred^{\mathcal{V}} - T_{\mathcal{U}\mathcal{V}} \cdot \gpred^{\mathcal{U}} \cdot \left(T_{\mathcal{U}\mathcal{V}}\right)^\dagger \right|\right|_n \; , \; 
\end{equation}
where $(T_{\mathcal{U}\mathcal{V}})_{\mu}^{\nu} = \partial v^\nu/ \partial u^\mu$  are the transition matrices between patches $\mathcal{U}$ and $\mathcal{V}$, with local coordinates $u$ and $v$, respectively. Moreover, $d$ is the number of patch transitions, and the default for the norm is $n=1$.
\item {\it K\"ahler class loss} $\cL_{\text{Kclass}}$\\
Finally, since we would like to compute the Ricci-flat metric for given K\"ahler parameters $t_\alpha$, we require a loss contribution which enforces the correct K\"ahler class\footnote{This is not essential for manifolds with $h^{1,1}(X)=1$ (such as the quintic), since all K\"ahler classes are obtained by simple metric rescaling. However, it is critical in order to generate metrics with a well-defined K\"ahler class for $h^{1,1}(X)>1$.}. It may appear this is not necessary for the $\phi$-model --- provided $g_\FS$ is set up with the correct K\"ahler class, the addition $\partial\bar{\partial}\phi$ is exact and, hence, does not change the class. While this is true mathematically, it depends on $\phi$ being a function (rather than a section), a condition which is difficult to enforce on a NN. For the other models in Table~\ref{tab:ansatz} there is no mathematical reason for the K\"ahler class to remain unchanged. In order to fix the K\"ahler class we define the {\it K\"ahler class loss function} as
\begin{equation} \label{eq:tvolkloss}
\cL_{\text{Kclass}} = \frac{1}{h^{1,1}(X)}\sum_{\alpha=1}^{h^{1,1}(X)}\left|\left|\mu_{t}({\cal O}_X(e_\alpha))-\int_X J_\text{pr}^2\wedge F_{\FS,\alpha}\right|\right|_n \; .
\end{equation}
Here, ${\cal O}_X(e_\alpha)$ are the line bundles specified by integers $(e_\alpha)^\beta=\delta_\alpha^\beta$, $F_{\FS,\alpha}=-i J_\alpha/(2\pi)$ are their associated field strengths and $\mu_t({\cal O}_X(e_\alpha)$ their slopes for the given K\"ahler parameters $t_\alpha$, as computed from the topological formula, Eq.~\eqref{eq:slopeizec}.   For the norm, the default is $n=1$. Conversely, these slope values on the ``line bundle basis'' ${\cal O}_X(e_\alpha)$, $\alpha=1,\ldots ,h^{1,1}(X)$, fix the K\"ahler parameters uniquely and this is the basic idea behind  Eq.~\eqref{eq:tvolkloss}: by enforcing the correct slopes we enforce the correct K\"ahler class. 

The slope integral in Eq.~\eqref{eq:tvolkloss} is evaluated numerically using MC integration, that is, integrals of the form~\eqref{eq:slope} are computed using Eq.~\eqref{intf2}. Note that this integral requires a large batch size to provide a good approximation to the slope, while the other losses benefit from gradient descent with small mini-batches. Hence, we break each training epoch of the NN into two optimization steps with two different batch sizes (we do use the same optimizer, to transfer some of the information, like momentum, between the two batches). A small batch size (default is 64 points) is used for all above contributions to the loss function while a large one (defaults to $10,000$ points or the entire data set, whichever is smaller) is used for the K\"ahler class loss. We have observed that also keeping the sigma loss in the updates with the large batch size further improves stability of the code. 
\end{itemize}
For all these loss functions, the derivatives with respect to the input coordinates are computed with TensorFlow's automatic differentiation, which works reliable for all NNs. 
 
A few comments are in order. By virtue of the Calabi-Yau theorem, the Monge-Amp\`ere loss $\cL_{\text{MA}}$ is sufficient to enforce Ricci-flatness of the metric. The Ricci-loss $\cL_{\text{Ricci}}$ has been included as a potential additional check but, since its computation is costly (as it involves taking two further derivatives), it is disabled by default. Moreover, for the $\phi$-model which is K\"ahler by construction, the K\"ahler loss $\cL_{\text{dJ}}$ is disabled. Similarly, the transition loss is disabled for this network (this is guaranteed as long as $\phi$ is a function). Disabling these losses for the $\phi$-model shortens training time and has had no ill effects on performance.

After the NN is trained, we require an error measure that tells us how well the NN approximates the quantity we are learning, and how well it compares with other approximation schemes. For our purposes, we consider error measures derived from the Monge-Amp\`ere equation and the requirement of Ricci-flatness. There are two established benchmark measures in the literature~\cite{Douglas:2006hz,Anderson:2010ke}, the sigma and Ricci measures, $(\sigma,\mathcal{R})$. The $\sigma$ measure is obtained by integrating, using the MC approximation as a weighted sum, the Monge-Amp\`ere equation 
\begin{equation}
\sigma = \frac{1}{{\rm Vol}_{\Omega}} \int_X \left| 1- \kappa\; \frac{\Omega \wedge \overline{\Omega}}{(J_\text{pr})^n}\right|\; ,
\end{equation}
where  ${\rm Vol}_{\Omega} = \int \text{d}\vO$. The Ricci measure results from integrating the Ricci scalar
\begin{equation}
\mathcal{R} = \frac{{\rm Vol}_{\CY}^{1/3}}{{\rm Vol}_{\Omega}} \int_X |R_\text{pr}| \; .
\end{equation}
It should be noted that a small Ricci measure is necessary, but not sufficient, to guarantee that the entire Ricci tensor is small. One could explore other (integrated) curvature measures, formed from other contractions of the Ricci tensor, which, however, we have not done in this work. We also provide measures to monitor the quality of the K\"ahler, transition, and volume approximations.

%%%%%%%%%%%%%%%%%%%%%%%%%%%%%%%%%%%%%%%%%%%%%%%%%%%%%%%%%%%%%%%%%%%%%%%%%%%%
\section{Results}
\label{sec:results}

In this section, we report the results from experiments where CY metrics are learned on CICY and KS manifolds.  These runs illustrate that the ML routines that we have developed perform well on CY manifolds from these lists.  As we will see,  there is some variance in the performance of the different NNs. The $\phi$ network emerges as the most efficient learner, and the small values reported on the $\sigma$ and $R$ error measures corroborate the accuracy of the obtained metrics. We furthermore find that the point-generating routines are efficient with run-times ranging from seconds (for the quintic) to an hour (for the KS CY example) on a CPU of a standard laptop. We expect this efficiency will persist on CYs of similar complexity.  For ease of notation, we will henceforth refer to the approximate CY metric obtained at the end of training as $g_\CY$.

\subsection{Quintic}
\label{sec:quintic}
The Quintic three-fold is specified as the vanishing locus of a quintic polynomial in the projective ambient space $\mathcal{A} = \mathbb{P}^4$, and appears in both the CICY and Kreuzer-Skarke lists. It has one K\"ahler modulus, which measures the overall volume of the manifold, and 101 complex structure moduli. A particular, simple instance of this family of manifolds is the Fermat quintic, given by the vanishing locus of  $Q = \sum_{i=0}^4 z_i^5$. The Fermat quintic is arguably the simplest CY three-fold, and is a benchmark model for numerical computations of CY metrics, whose metric has been approximated using a variety of approaches~\cite{donaldson2005numerical,Douglas:2006rr,Braun:2007sn,Braun:2008jp,Douglas:2006hz,Headrick:2009jz,Anderson:2020hux,Jejjala:2020wcc,Afkhami-Jeddi:2021qkf}. Given its simplicity, the Fermat quintic provides an excellent setting for comparing the performance of the different metric Ans\"atze in Table \ref{tab:ansatz}, and this is the main purpose of this subsection. 

\subsubsection{Comparison of  Ans\"atze for CY metric predictions}\label{sec:qmetric}

For these experiments, we  have generated a  random point sample of $198,000$ points on the quintic using the CICY point generator of {\texttt cymetric}. For each metric Ansatz in Table~\ref{tab:ansatz}, we have constructed  a fully connected feed-forward network with three hidden layers, each containing 64 nodes, and GELU activation functions.\footnote{Before these experiments, we have performed hyperparameter tuning via box searches for some of the hyperparameters, including for the number of hidden layers and the type of activation function.} These have been trained using an Adam optimizer on the point sample, with a $0.1$ validation split of the data. The NNs were initialized with weights sampled from a Gaussian distribution. For the networks that learn a correction to the reference metric $\gFSX$, we select a Gaussian distribution $\mathcal{N}(0,\;0.01)$ with small variance, whereas the free NN is initialized with $\mathcal{N}(0,\;1)$. 

\begin{figure}[t]
		\centering
		\includegraphics[width=1.\textwidth]{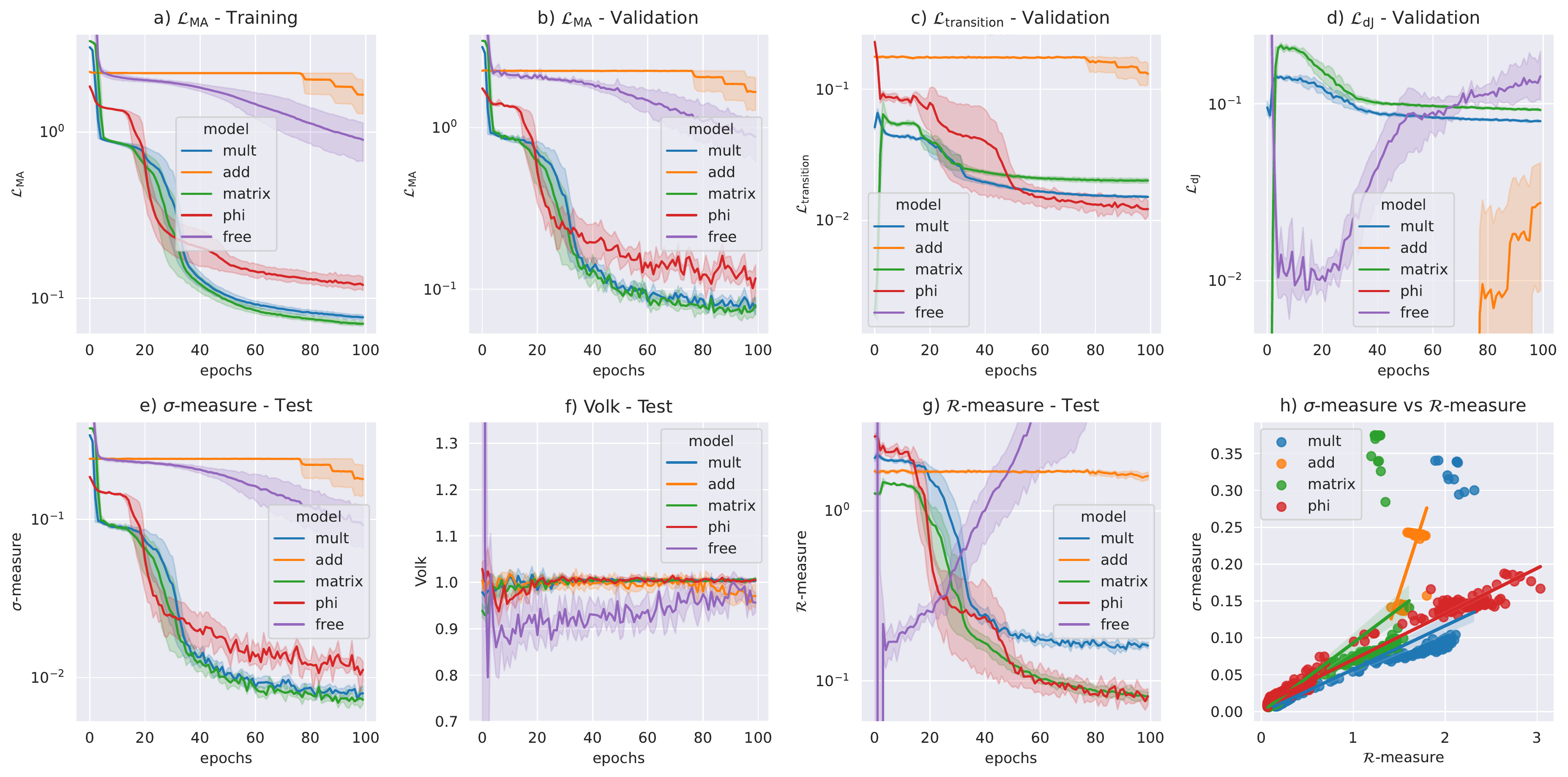}
		\caption{Fermat Quintic experiments: a) Monge-Amp\`ere loss on training data;  b)+c)+d) Monge-Amp\`ere, transition and K\"ahler loss on validation data; e) $\sigma$-measure f) volume  and g) $\mathcal{R}$-measure on test data; h) the linear relationship between improvement in $\sigma$-measure and $\mathcal{R}$-measure. The plots show the averaged performance of five separate experiments for each model, including 95\% confidence intervals as light-hue bands around each curve. \label{fig:fermat}}
	\end{figure}

For each metric Ansatz, five experiments over 100 epochs were performed. The results of these experiments are shown in Figure~\ref{fig:fermat} (and were first published in the 2021 NeurIPS conference, \cite{Larfors:2021pbb}). Note that for these quintic runs, the volume of the manifold has been normalized to 1. As shown in subplot f), all networks preserve this value of the volume.

%Plots

It is evident from Figure~\ref{fig:fermat} that learning takes place for all network Ans\"atze.  It is also clear that the performance of the NNs differ: the multiplication and $\phi$ models outperform the additive and free networks. In fact, the plots show that the training does not complete for the latter models: even after 100 epochs, most losses and error measures have not reached a stable value. In contrast, the two multiplication models, and the $\phi$ network, reduce the relevant losses and errors by approximately two orders of magnitude, and obtain stable values after the 100 epoch experiments are completed. 

In more detail, we see that the Monge-Amp\`ere loss, shown in subplots a) and b) for training and validation data, respectively, is learned perfectly by the best performing networks.
The similarity in performance between training and validation set implies that overfitting is not a problem. The transition loss on validation data is presented in subplot c). Again, learning is evident for all networks but for the additive NN the loss function only starts decreasing after 80 epochs. The final plot on the first row of Figure~\ref{fig:fermat} shows the  K\"ahler  loss on validation data. The $\phi$ model does not appear in this plot, since it provides a K\"ahler metric by default (as long as $\phi$ is a function, which is guaranteed by the preservation of the K\"ahler class, which is automatic up to a scale factor for the Quintic). Interestingly, the multiplicative models perform worse than the additive model on this loss. This is to be expected: multiplying the metric with a non-constant matrix will generically result in an associated 2-form which fails to close. On the other hand, for the additive network, while the addition to the metric will generically break the K\"ahler constraint, the smallness of the addition implies that this breaking will be small. One should expect the result to reflect this, and it is reassuring that it does.

On the second row of Figure~\ref{fig:fermat} it is checked that the volume of the manifold is not changed during learning. We also compute the $\sigma$ and $\mathcal{R}$ error measures. These checks are performed on a separate test set of $22,000$ points. The $\sigma$ and  $\mathcal{R}$ measures are benchmarks for the quality of CY metric approximations, and their low values testify to the performance of the NNs. The $\phi$-models with lowest validation loss reach, within an hour on a single CPU of a standard laptop, a mean accuracy of $\sigma = 0.0086$ and $\mathcal{R} = 0.076$ and thus matches $k=20$ in Donaldson algorithm~\cite{Anderson:2010ke} 
with a training time of 35 years on $4.6 \cdot 10^8$ points~\cite{Ashmore:2019wzb}. On the other hand, the energy functional methods obtain a similar accuracy already at $k=4$ \cite{Headrick:2009jz} and other machine learning methods report similar runtimes as ours.\footnote{We recall that reducing runtime for ML experiments is not the primary goal of the {\texttt cymetric} package; rather the benefit of the new package is that it can be applied to a wider range of CY manifolds than previously possible and that it computes the Ricci-flat metric for a given K\"ahler class.}

To the best of our knowledge, we are the first to systematically study $\mathcal{R}$, and to demonstrate, in subplot h), a linear relation, $\sigma \approx 0.06 \mathcal{R}$ for the $\phi$-model, between optimization of the surrogate Monge-Amp\`ere equation~\eqref{eq:MA} and decrease in Ricci measure $\mathcal{R}$. While this is to be expected from the CY theorem, a numerical routine could in principle deviate from this result. It is reassuring that our implementation does not and that there is no tension between these two constraints.

Another conclusion from these experiments is that the $\phi$-model performs better than the other models, at least for simple CY three-folds.\footnote{This conclusion is also corroborated by experiments reported on in Ref.~\cite{Schneider:2022ssn}.} While this is a result that should  be treated with some caution, we have reached similar conclusions on other manifolds, and will thus focus on this model in the following sections.

\subsection{Bi-cubic}\label{sec:bicubic}
In this sub-section we describe our numerical results for the bi-cubic CY, a manifold with Picard number two. Specifically, we calculate the numerical Ricci-flat CY metrics at various specific points in the bi-cubic K\"ahler moduli space. For these metrics we perform two basic checks to confirm that the correct point in K\"ahler moduli space has indeed been reached: a numerical calculation of the CY volume and a numerical calculation of various line bundle slopes, each of which can be compared with the exact result obtained from a topological formula. As an example application, we also compute the HYM connection for a specific line bundle on the bi-cubic numerically.

\subsubsection{Basic set-up}
The bi-cubic CY is defined as a hypersurface in the ambient space ${\cal A}=\mathbb{P}^2\times\mathbb{P}^2$ and it is contained in both the CICY and the KS lists. Specifically, it is given as the zero locus in ${\cal A}$ of a bi-degree $(3,3)$ polynomial. Its Hodge numbers are $(h^{1,1}(X),h^{2,1}(X))=(2,83)$ and its non-zero intersection numbers are $d_{112}=d_{122}=3$. Throughout our calculations, we fix a specific point in complex structure moduli space by focusing on the defining polynomial
\begin{align}
p=&0.96 x_0^3 y_0^3+0.79 x_0^3 y_1^3+0.49 x_0^3 y_2^3+0.51 x_0^3 y_0 y_1 y_2+0.05 x_1 x_0^2
   y_0 y_1^2+0.99 x_2 x_0^2 y_0 y_2^2+0.81 x_1 x_0^2 y_1 y_2^2\nonumber\\
   &+0.39 x_2 x_0^2 y_0^2 y_1+0.54 x_1 x_0^2 y_0^2 y_2+0.87 x_2 x_0^2 y_1^2 y_2+0.62 x_1 x_2 x_0 y_0^3+0.62 x_1
   x_2 x_0 y_1^3+0.62 x_1 x_2 x_0 y_2^3\nonumber\\
   &+0.81 x_2^2 x_0 y_0 y_1^2+0.87 x_1^2 x_0 y_0 y_2^2+0.54 x_2^2 x_0 y_1 y_2^2+0.99 x_1^2 x_0 y_0^2 y_1+
      0.05 x_2^2 x_0 y_0^2 y_2+0.39 x_1^2 x_0 y_1^2 y_2\nonumber\\
   &+0.87 x_1 x_2 x_0 y_0 y_1 y_2+0.49 x_1^3 y_0^3+0.79 x_2^3
   y_0^3+0.96 x_1^3 y_1^3+0.49 x_2^3 y_1^3+0.79 x_1^3 y_2^3+0.96 x_2^3 y_2^3\nonumber\\
   &+0.54 x_1^2x_2 y_0 y_1^2+0.39 x_1 x_2^2 y_0 y_2^2+0.05 x_1^2 x_2 y_1 y_2^2+0.87 x_1 x_2^2 y_0^2
   y_1+0.81 x_1^2 x_2 y_0^2 y_2+0.99 x_1 x_2^2 y_1^2 y_2\nonumber\\
   &+0.51 x_1^3 y_0 y_1 y_2+0.51x_2^3 y_0 y_1 y_2\; ,
\end{align}
where $(x_0,x_1,x_2)$ and $(y_0,y_1,y_2)$ are homogeneous coordinates on the two $\mathbb{P}^2$ factors. We will calculate the Ricci-flat metric for the seven points in K\"ahler moduli space indicated in Table~\ref{tab:bicubickahler}.
\begin{table}[t]
\centering
\begin{tabular}{|c||c|c|c|c|c|c|c|}\hline
case&1&2&3&4&5&6&7\\\hline\hline
$t_{(i)}$
&
$\left(
\begin{array}{c}
 1.414 \\
 1.414 \\
\end{array}
\right)$
&
$\left(
\begin{array}{c}
 0.687 \\
 1.878 \\
\end{array}
\right)$
&
$\left(
\begin{array}{c}
 0.421 \\
 1.955 \\
\end{array}
\right)$
&
$\left(
\begin{array}{c}
 0.299 \\
 1.977 \\
\end{array}
\right)$
&
$\left(
\begin{array}{c}
 0.962 \\
 1.753 \\
\end{array}
\right)$
&
$\left(
\begin{array}{c}
 1.092 \\
 1.676 \\
\end{array}
\right)$
&
$\left(
\begin{array}{c}
 0.853 \\
 1.809 \\
\end{array}
\right)$\\[4mm]\hline
${\cal O}_X(k_{(i)})$&${\cal O}_X(1, -1)$&${\cal O}_X(1, -2)$&${\cal O}_X(1, -3)$&${\cal O}_X(1, -4)$&${\cal O}_X(2, -3)$&${\cal O}_X(3, -4)$&${\cal O}_X(3, -5)$\\\hline
\end{tabular}
\caption{Choices $t_{(i)}$, where $i=1,\ldots , 7$, of the K\"ahler parameters for the bi-cubic and corresponding slope zero line bundles with line bundle integers $k_{(i)}$.}\label{tab:bicubickahler} 
\end{table}
For each of these choices of $t$ there is a corresponding line bundle ${\cal O}_X(k)$ with slope zero which is given in the last row of Table~\ref{tab:bicubickahler}.

\subsubsection{Point sampling and training with $\phi$-model}
For each of the seven choices of K\"ahler parameters in Table~\ref{tab:bicubickahler}, we have generated $100,000$  points (with 0.1 validation split) on the bi-cubic, using the Mathematica point generator of the \texttt{cymetric} package. For the $\phi$-model and a neural network with width $64$, depth $3$, GELU activation functions and initialization with $\mathcal{N}(0,\;0.01)$, training has been carried out for $100$ epochs, using the Adam optimizer with a batch size of $64$ and a learning rate of $1/1000$.  Training has been completed on a CPU of a standard laptop in about three hours, for each of the seven choices of K\"ahler parameters.

\begin{figure}
\centering
\includegraphics[width=0.3\textwidth]{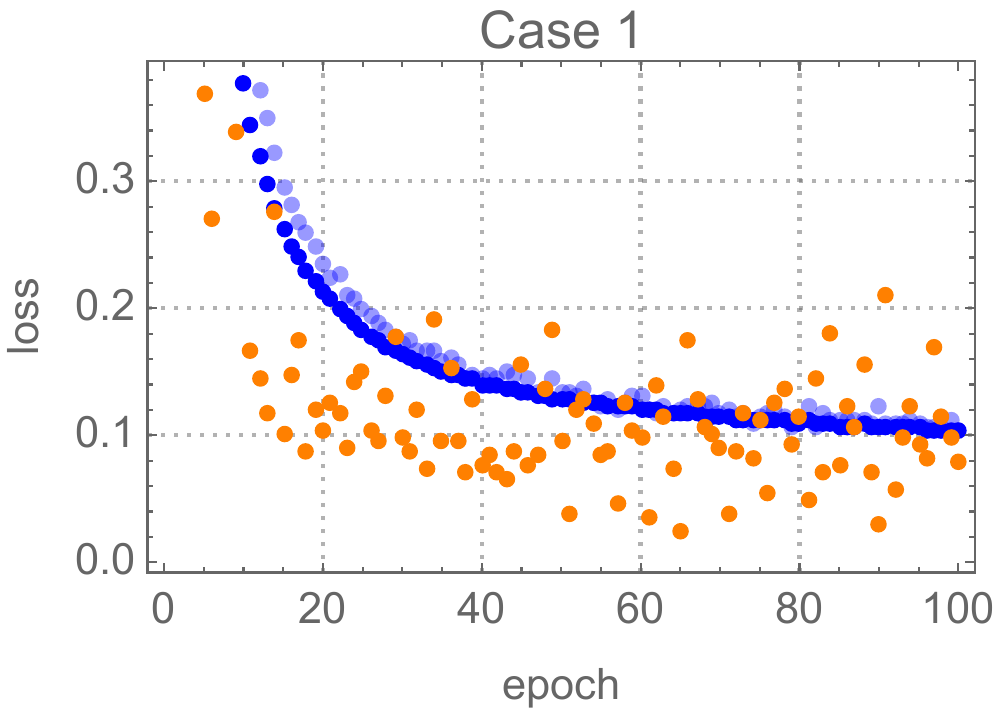}\hskip 5mm
\includegraphics[width=0.3\textwidth]{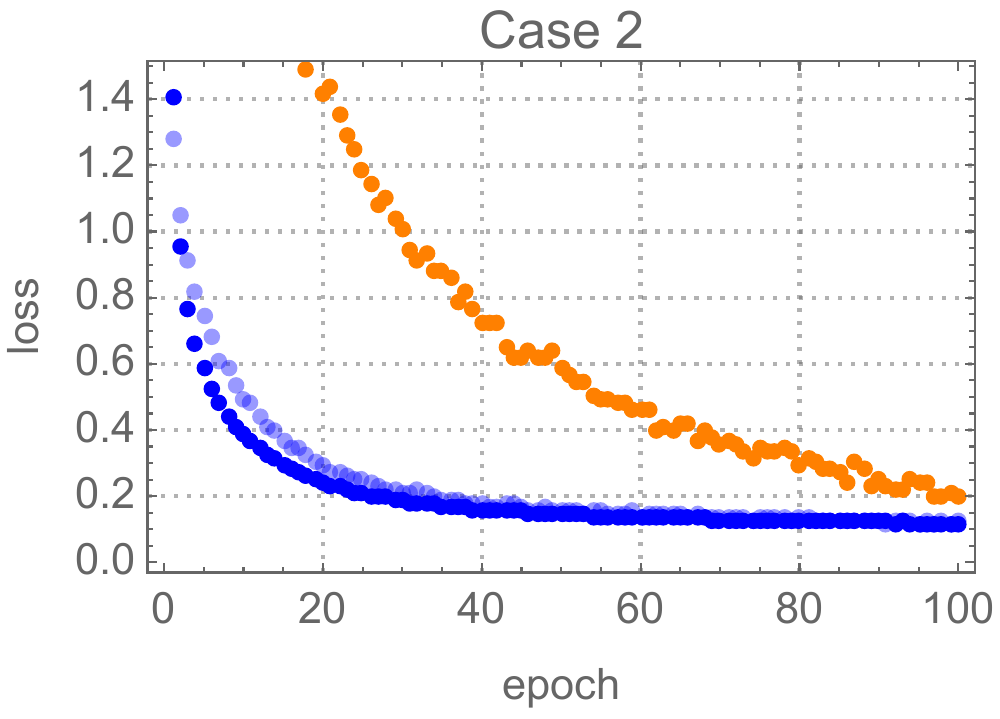}\hskip 5mm
\includegraphics[width=0.3\textwidth]{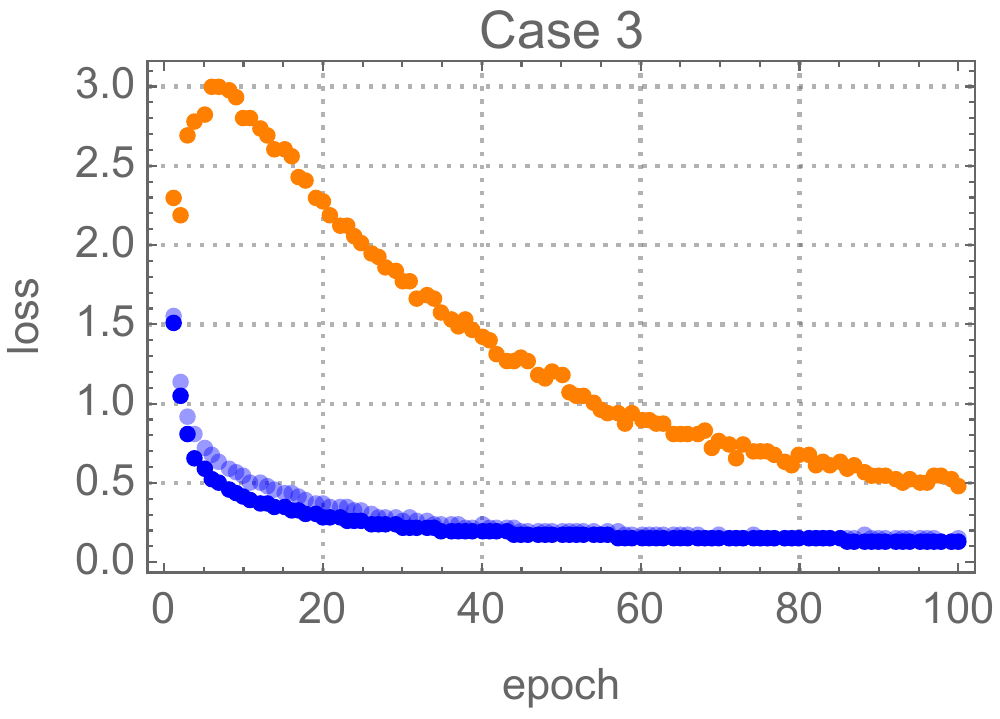}\\[2mm]
\includegraphics[width=0.3\textwidth]{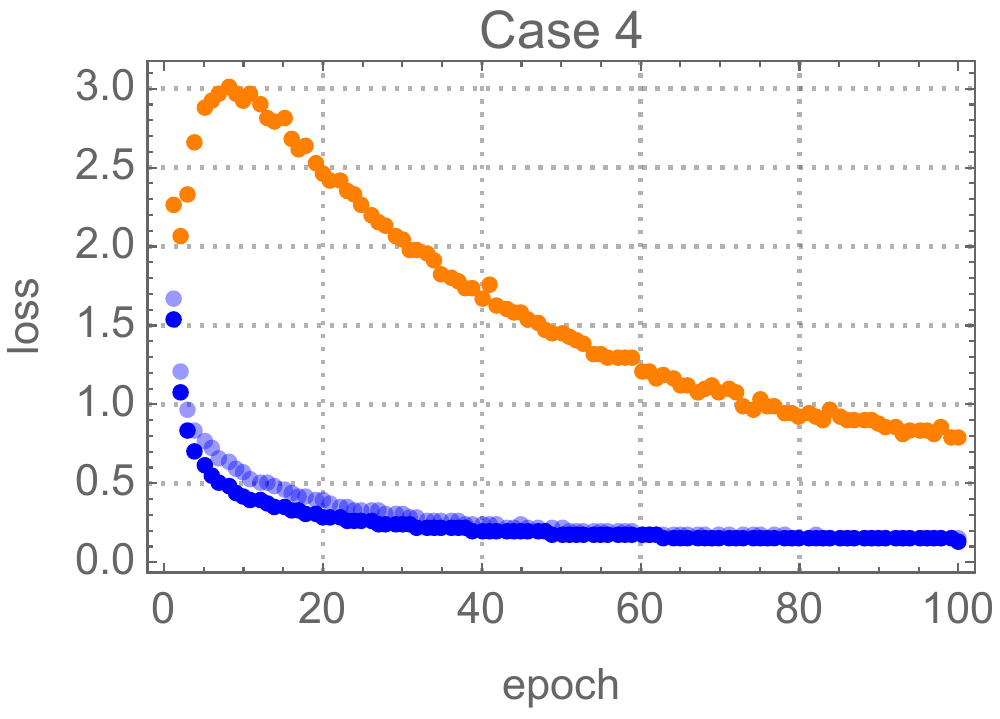}\hskip 5mm
\includegraphics[width=0.3\textwidth]{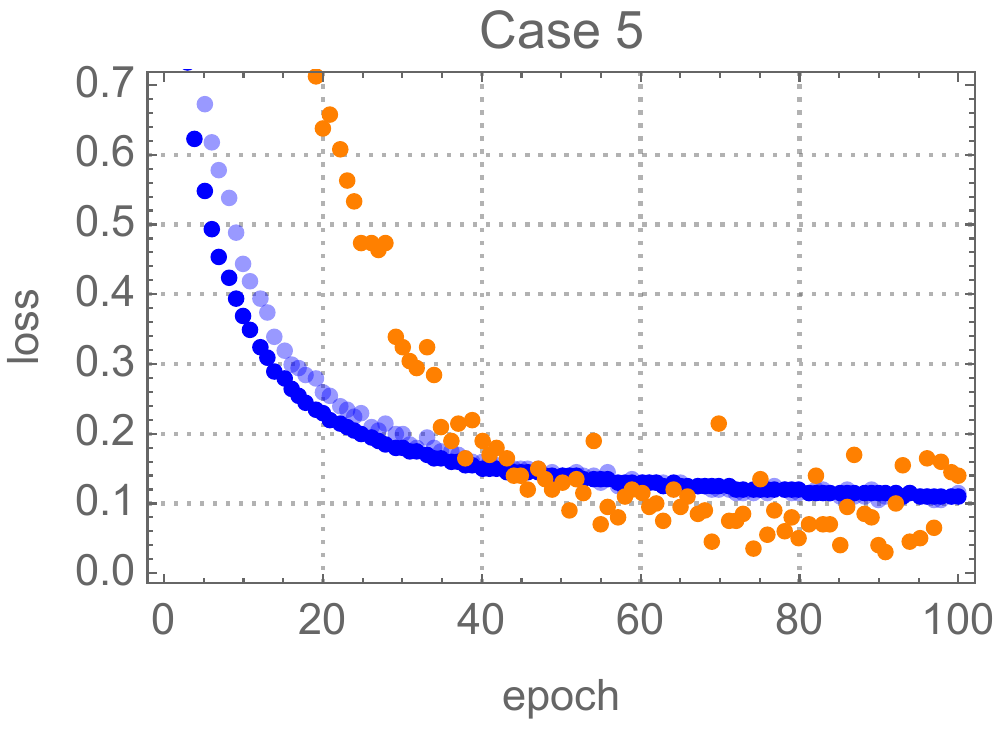}\hskip 5mm
\includegraphics[width=0.3\textwidth]{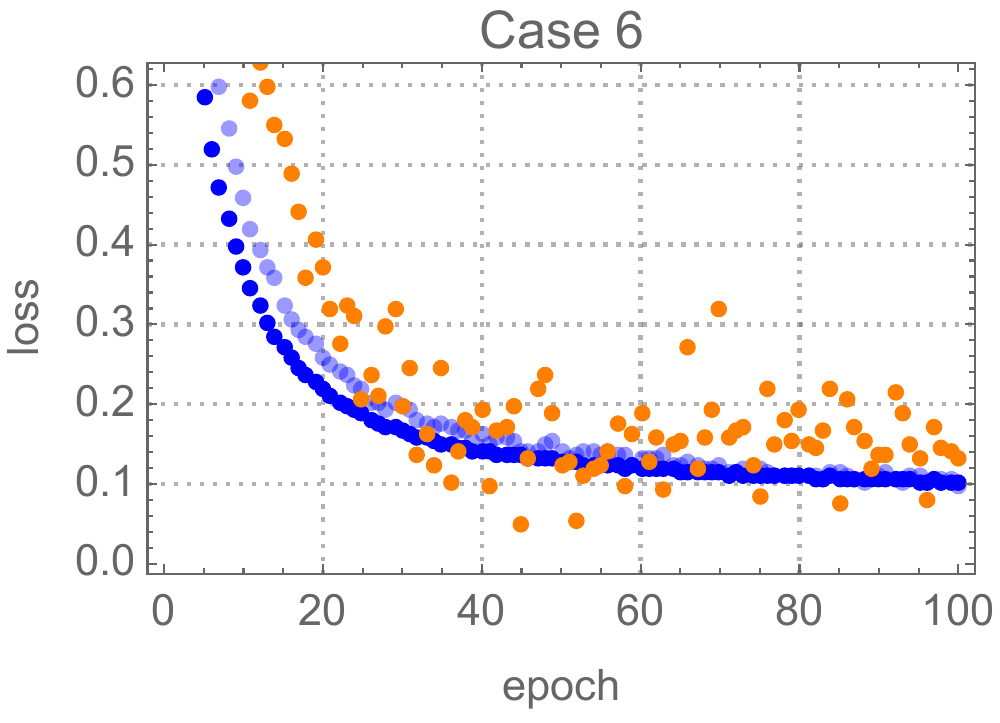}\\[2mm]
\includegraphics[width=0.3\textwidth]{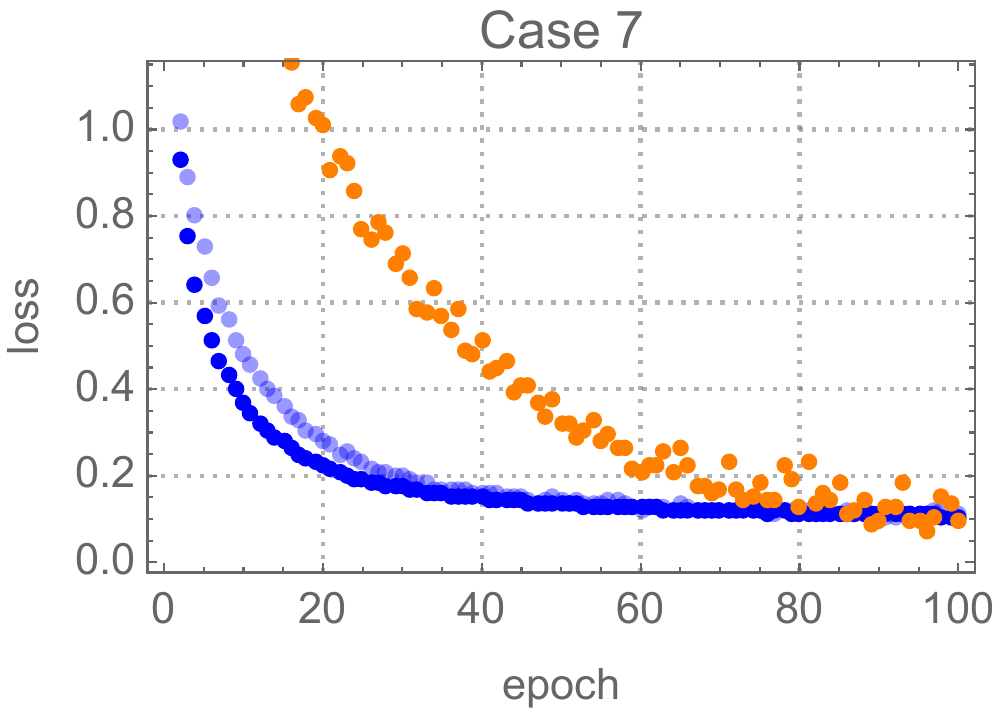}\hskip 5mm
\includegraphics[width=0.3\textwidth]{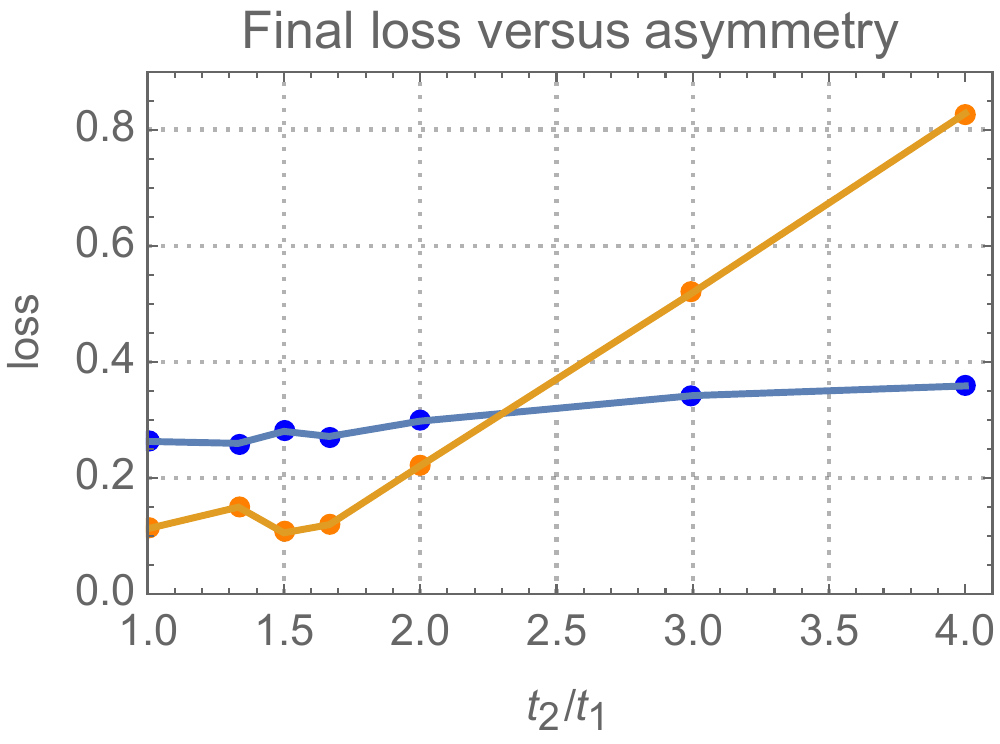}
\caption{Bi-cubic training curves for the seven choices of K\"ahler parameters in Table~\ref{tab:bicubickahler}. The last plot represents the final loss, obtained by averaging over the last $10$ epochs, as a function of $t^2/t^1$\\ (orange:  $\cL_{\text{Kclass}}$, blue: $4\times \cL_{\text{MA}}$, both on training data, light-blue: $4\times \sigma$ measure on validation data). }\label{fig:bicubicloss}
\end{figure}

The only two relevant contributions to the loss are the Monge-Amp\`ere loss and the K\"ahler class loss which are shown, together with the $\sigma$ error measure in Fig.~\ref{fig:bicubicloss}. 
Evidently, training is efficient and successful for all seven cases. The last plot in Fig.~\ref{fig:bicubicloss} shows the final MA and Kclass loss, obtained by averaging over the last $10$ epochs, as a function of the modulus ratio $t^2/t^1$, which can be seen as a measure of the asymmetry of the manifold. There is a clear tendency for the final loss to increase with increasing asymmetry, a behavior which is intuitively expected.

\subsubsection{Volume computations}
From the runs described above we have obtained numerical results for the Ricci-flat CY metric $g_\CY(p_i)$ at $100,000$ points $p_i$ on the bi-cubic and for seven choices of K\"ahler parameters. In addition, the \texttt{cymetric} package provides the weights $w_i$, the auxiliary weights $\tilde{w}_i$ and the Fubini-Study metric $g_\FS(p_i)$ at those points.  

As a first check, we would like to compute the CY volume for all seven cases, based on Eq.~\eqref{CYvol}. The exact results are obtained from the intersection formula on the RHS of Eq.~\eqref{CYvol}. Alternatively, these volumes can be computed by integrating over the CY or the FS measure, as in the middle of Eq.~\eqref{CYvol}, and we will evaluate these integrals numerically, as explained in Section~\ref{sec:torsamp}. Explicitly, we compute for all seven cases
\begin{equation} \label{eq:volumes}
V_{\rm int}=\frac{1}{6}d_{\alpha\beta\gamma}t^\alpha t^\beta t^\gamma \; , \;  V_\FS= \frac{1}{N}\sum_{i=1}^N\tilde{w}_i\, {\rm det}(g_\FS(p_i)) \; , \; V_{\CY}= \frac{1}{N}\sum_{i=1}^N\tilde{w}_i\, {\rm det}(g_\CY(p_i)) \; ,
\end{equation}
where we recall that $g_\CY$ refers to the network's prediction for the CY metric after the training is completed. The results are given in Table~\ref{tab:bicubicvolumes}.
\begin{table}[t]
\centering
\begin{tabular}{|c||c|c|c|c|c|c|c|}\hline
case&1&2&3&4&5&6&7\\\hline\hline
$V_{\rm int}$&8.49&4.97&2.93&2.02&6.87&7.59&6.16\\\hline
$V_{\FS}$&8.49&4.50&2.94&2.03&6.91&7.58&6.26\\
error&$<1\%$&$<1\%$&$< 1\%$&$< 1\%$&$< 1\%$&$<1\%$&$\sim 2\%$\\\hline
$V_{\CY}$ &8.56&5.03&2.96& 2.03&6.86&7.58&6.28\\
error&$< 1\%$&$\sim 1\%$&$< 1\%$&$< 1\%$&$< 1\%$&$< 1\%$&$\sim 2\%$\\\hline
\end{tabular}
\caption{Exact volume from intersection form (row 2), and volume from numerical integration with $g_\FS$ (row 3) and $g_\CY$ (row 4), for the seven choices of K\"ahler parameters in Table~\ref{tab:bicubickahler}.}\label{tab:bicubicvolumes} 
\end{table}

The volumes computed with the Fubini-Study metric are in good agreement with the exact results, with most errors at the level of $1\%$ or below. In this case, the only uncertainty comes from point sampling (as the Fubini-Study metric is known exactly) so these results confirm that our method for sampling points works and that $100,000$ points are sufficient. 

The more important results in Table~\ref{tab:bicubicvolumes} are the volumes computed with the Ricci-flat CY metric, given in the last row. Again, the accuracy is impressive at $1\%$ or less for most cases. Overall, these results confirm that the Ricci-flat CY metrics obtained are indeed in the prescribed K\"ahler class -- only then can the volume be expected to come out correctly. We emphasize that obtaining the correct volume has \emph{not} been built into the loss function. Rather, the correct K\"ahler class has been enforced during training by imposing the K\"ahler class loss~\eqref{eq:tvolkloss}. Table~\ref{tab:bicubicvolumes} provides strong evidence that this method does indeed work.

\subsubsection{Slope computations}
While the correct slope has been imposed during training it is still worth checking that our numerical metrics can be used for accurate slope computations. For the seven values of K\"ahler parameters $t_{(i)}$ and the seven line bundles ${\cal O}_X(k_{(j)})$ in Table~\ref{tab:bicubickahler} we can compute a $7\times 7$ slope matrix with entries $\mu_{t_{(i)}}({\cal O}_X(k_{(j)}))$. The accurate entries of this matrix are obtained from the topological formula~\eqref{eq:slopeizec} and are given by
\begin{equation}\label{eq:slopesbicubicizec}
\left[
\begin{array}{rrrrrrr}
 0.00 & -18.00 & -36.00 & -54.00 & -18.00 & -18.00 & -36.00 \\
 9.16 & 0.00 & -9.16 & -18.33 & 9.16 & 18.33 & 9.16 \\
 10.94 & 5.47 & 0.00 & -5.47 & 16.41 & 27.34 & 21.87 \\
 11.46 & 7.64 & 3.82 & 0.00 & 19.1 & 30.57 & 26.75 \\
 6.45 & -6.45 & -19.34 & -32.24 & 0.00 & 6.45 & -6.45 \\
 4.85 & -9.7 & -24.25 & -38.8 & -4.85 & 0.00 & -14.55 \\
 7.63 & -3.82 & -15.26 & -26.71 & 3.82 & 11.45 & 0.00 \\
\end{array}
\right]
\end{equation}
where the rows are labeled by the values of the K\"ahler parameters and the columns by the line bundles. The zeros along the diagonal indicate the existence of HYM connections for these cases, that is, for the line bundles ${\cal O}_X(k_{(i)})$ and their associated K\"ahler parameters $t_{(i)}$.

Alternatively, we can work out the slope by carrying out the integrals in Eq.~\eqref{eq:slope} numerically. Doing this first for the FS metrics, that is, evaluating $\frac{2}{N\pi}\sum_{i=1}^N\tilde{w}_i\, {\rm det}(g_\FS(p_i))\,\rho_\FS(p_i)$ gives
\begin{equation}
\left[
\begin{array}{rrrrrrr}
 0.11 & -17.84 & -35.78 & -53.73 & -17.73 & -17.62 & -35.57 \\
 9.12 & -0.13 & -9.37 & -18.62 & 8.99 & 18.11 & 8.86 \\
 10.95 & 5.46 & -0.03 & -5.52 & 16.41 & 27.35 & 21.86 \\
 11.47 & 7.65 & 3.82 & -0.01 & 19.12 & 30.6 & 26.77 \\
 6.27 & -6.79 & -19.84 & -32.9 & -0.52 & 5.75 & -7.31 \\
 4.95 & -9.53 & -24.01 & -38.49 & -4.58 & 0.37 & -14.11 \\
 7.34 & -4.43 & -16.19 & -27.95 & 2.91 & 10.25 & -1.51 \\
\end{array}
\right]
\end{equation}
Agreement with the exact results in \eqref{eq:slopesbicubicizec} is impressive, with an average error of the non-zero slope of $\sim 3\%$, suggesting that the point sample size is sufficient, at least for the purpose of computing slopes. Using the Ricci-flat CY metrics and evaluating $\frac{2}{N\pi}\sum_{i=1}^N\tilde{w}_i\, {\rm det}(g_\CY(p_i))\,\rho_\CY(p_i)$, we find
\begin{equation}
\left[
\begin{array}{rrrrrrr}
 0.03 & -17.97 & -35.98 & -53.98 & -17.95 & -17.92 & -35.92 \\
 8.86 & -0.42 & -9.70 & -18.98 & 8.43 & 17.29 & 8.01 \\
 10.11 & 4.52 & -1.07 & -6.65 & 14.63 & 24.74 & 19.15 \\
 9.96 & 5.96 & 1.97 & -2.02 & 15.92 & 25.87 & 21.88 \\
 6.38 & -6.45 & -19.29 & -32.12 & -0.07 & 6.31 & -6.53 \\
 4.96 & -9.41 & -23.77 & -38.13 & -4.45 & 0.51 & -13.86 \\
 7.53 & -4.12 & -15.76 & -27.4 & 3.41 & 10.93 & -0.71 \\
\end{array}
\right]
\end{equation}
Again, this is in good agreement with the exact results in \eqref{eq:slopesbicubicizec}, with an average error of the non-zero slopes of $\sim 7\%$, and it confirms that we have obtained Ricci-flat CY metrics with the correct, intended K\"ahler class. We also note that, in most cases, the results are accurate enough to distinguish zero slope cases  (along the diagonal) which allow for a HYM connection from cases with non-zero slope. For highly asymmetric manifolds, this distinction is less clear, which is related to the fact that training takes longer in these cases.

\subsubsection{HYM connection}
As a final application of our results we would like to compute an approximation to the HYM connection on a line bundle over the bi-cubic, following the procedure described in Section~\ref{sec:hym}. To do this we focus on the line bundle ${\cal O}_X(1,-1)$ and K\"ahler parameters $t_1=t_2\simeq 1.414$, corresponding to case 1 in Table~\ref{tab:bicubickahler}. As the result in \eqref{eq:slopesbicubicizec} (specifically, the $(11)$ entry of this matrix) shows, the slope of ${\cal O}_X(1,-1)$ vanishes for this choice of K\"ahler parameters, so that a HYM connection does indeed exist. 

Our method requires choosing a set of function~\eqref{eq:fbasis} in which to expand the various quantities. We do this using the sections $\Gamma({\cal O}_X(1,1))$, a nine-dimensional space with basis $(x_a y_b)$, where $a,b=0,1,2$. This leads to $81$ functions of the form~\eqref{eq:fbasis}, explicitly given by
\begin{equation}\label{eq:fbicubic}
 f_{(ab)(cd)}=\frac{x_ay_b\bar{x}_c\bar{y}_d}{(|x_0|^2+|x_1|^2+|x_2|^2)(|y_0|^2+|y_1|^2+|y_2|^2)}\; .
\end{equation}
Note, these expressions are indeed of homogeneity degree zero in each set of $\mathbb{P}^2$ coordinates and, therefore, constitute functions on the ambient space and, by restriction, on the bi-cubic CY.

With these functions we compute the $81\times 81$ matrix $\Delta$ which represents the Laplacian and the $81$-dimensional vector ${\boldsymbol\rho}$ which represents the inhomogeneity in Laplace's equations, by evaluating the matrix elements in Eq.~\eqref{matel} by numerical integration, using our point sample and numerical Ricci-flat metric for case 1. We can check that the matrix $\Delta$ obtained in this way has a one-dimensional kernel which corresponds to constant functions on the bi-cubic, as can be verified by taking the linear combination of the functions~\eqref{eq:fbicubic} with a vector in the kernel. As expected, the source vector ${\boldsymbol\rho}$ is not contained in the image ${\rm Im}(\Delta)$, but we can verify that $|{\boldsymbol\rho}^\perp|/|{\boldsymbol\rho}^\parallel|\sim 0.03$, so the component of ${\boldsymbol\rho}$ orthogonal to ${\rm Im}(\Delta)$ is small. Then we solve Eq.~\eqref{betaeqla2} to find an $81$-dimensional solution vector ${\boldsymbol\beta}$ and the linear combination of the functions~\eqref{eq:fbicubic} formed with this vector provides gives our approximation of the function $\beta$. Via Eq.~\eqref{eq:FCY}, this function $\beta$ then determines the approximation to the HYM bundle metric. For this result we compute the quantity~\eqref{eq:HYMacc} as a measure of the accuracy of the approximation. This comes out at $\sim 3\%$, indicating a reasonably accurate HYM connection. 

\subsection{Toric hypersurface CY with Picard rank 2}\label{sec:toricresults}

To test the \texttt{cymetric} package's ability to predict CY metrics on KS CY manifolds, we select a CY manifold with Picard rank two from the KS list. After specifying the K\"ahler moduli as $(t_1,t_2)=(1,1)$ and the complex structure moduli by a random assignment of the coefficient of the defining polynomial, we generate points on the manifold as specified in Section~\ref{sec:torsamp}. These points are then used in training the $\phi$-model to predict the CY metric. 

\subsubsection{Geometric set-up and point sampling}

The selected manifold has a toric ambient space with two K\"ahler moduli. The vertices\footnote{The reader unfamiliar with toric geometry may wish to consult Appendix~\ref{app:ToricGeometryIntro}.}  
\begin{align}\label{eq:toricvert}
v_0=\left(\begin{array}{r}
-1 \\ -1 \\ -1 \\ 0 \\  \end{array}\right),\quad
v_1=\begin{pmatrix}
 0 \\ 0 \\ 0\\ 1 \end{pmatrix},\quad
v_2=\begin{pmatrix}
 0 \\ 0 \\ 1\\ 0 \end{pmatrix},\quad
v_3=\begin{pmatrix}
 0 \\ 1 \\ 0 \\ 0 \end{pmatrix},\quad
v_4=\left(\begin{array}{r}
 2 \\ 0 \\ 0\\ -1 \end{array}\right),\quad
v_5=\begin{pmatrix}
1 \\ 0 \\ 0  \\ 0 \\  \end{pmatrix} \, ,
\end{align}
span a fan in the lattice $N$ which defines a $\mathbb{P}^1$-fibered toric ambient space over $\mathbb{P}^3$, ${\cal A}=\mathbb{P}^1 \rightarrow \mathbb{P}^3$, of Picard rank two.  Each vertex corresponds to a homogeneous coordinate $x^i \sim v_i$, and a divisor $D_i = \{x^i = 0\}$; indeed, the vertices $v_0, v_2, v_3,v_5$ can be seen to correspond to $\mathbbm{P}^3$ by projection onto the first three entries,  and vertices $v_1,v_4$ give rise to a $\mathbbm{P}^1$ by projection onto the last entry. There is a unique fine regular star triangulation of the corresponding polytope, giving eight top-dimensional cones spanned by four out of the six $v_i$. Thus, ${\cal A}$ can be covered by eight patches, and in each such patch there are four affine coordinates. 

We can use the methods detailed in Appendix~\ref{app:ToricGeometryIntro} to compute the sections of the anti-canonical bundle, whose linear combinations give a homogeneous polynomial that specifies a CY hypersurface $X$ in $\cal A$. For this example, there are 80 sections, and we select random coefficients, corresponding to a generic choice for the complex structure moduli of the CY. Rather than writing out this lengthy expression, we note that a different random choice leads to comparable performance for the metric model.

Following the method described in Section~\ref{sec:torsamp}, we sample points on $X$ using the toric point generator of \texttt{cymetric}. For this we note that the GLSM charges are
\begin{equation}\label{eq:glsm}
\left(\begin{array}{rrrrrr}
1 & 0 & 1 & 1 & 0 & 1\\
0 &  1& 0 & 0 & 1 & -2
\end{array}\right) \; .
\end{equation}
The Stanley-Reisner ideal is given by 
\begin{equation}
\text{SRI} = \langle x^1 x^4, x^0 x^2 x^3 x^5 \rangle \,,
\end{equation}
from which we can of course also infer the fibration structure of $\mathbb{P}^1$ over $\mathbb{P}^3$.

Using the GLSM charges we may read off the homogeneous scalings of the toric coordinates, and deduce the relations between the divisors:
\begin{equation}
D_0 \sim D_2 \sim D_3 \; , \; D_1 \sim D_4 \; , \; D_5 \sim D_0-2D_1 \; .
\end{equation}
We may thus choose $J_1 = D_1$ and $J_2 = D_3$ as generators for the K\"ahler cone, with non-vanishing triple intersection numbers given by
\begin{equation} \label{eq:toricintnum}
d_{122}=d_{212}=d_{221} = 4 \; , \; d_{222} = 8  \; .
\end{equation}

Finally, the sections of the K\"ahler cone generators are, for this example, 
\begin{align}
H^0(J_1)=(x_1,x_4)\,,\qquad H^0(J_2)=(x_0,x_2,x_3,x_1^2 x_5,x_4^2 x_5,x_1 x_4 x_5)\,,
\end{align}
(which by \eqref{eq:glsm}  all have the appropriate scaling).  This defines the morphisms  $\Phi_{1,2}$ into $\mathbbm{P}^1$ and $\mathbbm{P}^5$, respectively, and allow us to construct points on the CY as discussed in Section~\ref{sec:torsamp}.

\subsubsection{Training with $\phi$-model}

We have performed five experiments using the $\phi$-model and a neural network with width $256$, $3$ hidden layers, GELU activation functions and initialization with $\mathcal{N}(0,\;0.01)$. The network is trained for $100$ epochs with a training set of $200,000$ points, and error measures are computed on a validation set of $20,000$ points. We chose default values for $\alpha$ using a SGD optimizer with learning rate $5/10000$ and momentum $0.95$, and the batch sizes for the optimization steps were $(128,\;10000)$. It was observed in the design phase that the two updates with different batch sizes required to minimize the Monge-Am\`ere loss while staying in the reference K\"ahler class, sometimes lead to the determinant of the predicted metric turning negative in isolated points during training. To ameliorate the long term effects of this behavior, training was conducted with lower learning rate and a larger momentum compared to the bicubic runs. The training took a couple of hours on a single GPU.

%Plots
	\begin{figure}[t]
		\centering
		\includegraphics[width=\textwidth]{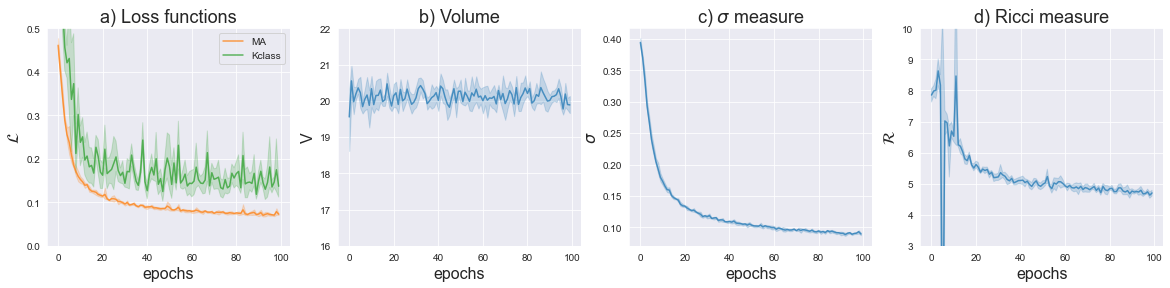}
		\caption{KS CY experiments: a) Monge-Amp\`ere and K\"ahler class loss on training  data;  b) volume c) $\sigma$ measure and d) $\mathcal{R}$ measure on validation data. The plots show the averaged performance of five separate experiments for the $\phi$ model, including 95\% confidence intervals as light-hue bands around each curve. \label{fig:ksmanifold}}
	\end{figure}

The results of the five experiments are shown in Figure~\ref{fig:ksmanifold}. We observe that the Monge-Amp\`ere loss, $\cL_{\text{MA}}$, decreases  steadily over training, totaling to a factor 4  decrease over 100 epochs. While this is certainly less impressive than the improvement on the quintic, learning clearly takes place. For the $\phi$ model, the transition and K\"ahler losses are automatically zero, and hence are disabled. The decrease in the K\"ahler class loss, $\cL_{\text{Kclass}}$, is also a factor 4, and the volume stays approximately constant around 20, which is the true value. The $\sigma$ measure, which is calculated on validation data, also decreases steadily over training, showing that the NN provides a better and better solution to the Monge-Amp\`ere equation. The most interesting part of this experiment is the evolution of the Ricci measure. During the initial 20 epochs, this goes through sharp spikes, corresponding to occurrences of points where the determinant of the predicted metric  $\gpred = g_\FS + \partial \bar{\partial} \phi$ becomes negative. This unphysical behavior is not lasting, and the training converges to a Ricci measure of around 5. While this is clearly non-zero, training decreases $\cal R$ by roughly $37\%$.

As for the bicubic, we can now compute the volume of the CY using the intersection numbers, FS and CY metric. We find, for $t_1=t_2=1$,
\begin{equation}
V_{\rm int}=20\; , \quad  V_\FS= 19.98\; , \quad V_{\CY}=  19.89 \; ,
\end{equation}
using Eq.~\eqref{eq:volumes} and the intersection numbers given in \eqref{eq:toricintnum}, and $V_\CY$ refers to the network's prediction for the CY volume after the training is completed, and is here computed as the mean of the five runs.\footnote{For the five experiments, we find $V_{\CY}=\{19.92 \, , 19.73 \, , 20.34 \, , 19.49 \, , 19.96 \}$, respectively.} We find that the volumes agree within 1 \%, a result that testifies to the accuracy of the point generating routine and metric prediction.

%%%%%%%%%%%%%%%%%%%%%%%%%%%%%%%%%%%%%%%%%%%%%%%%%%%%%%%%%%%%%%%%%%%%%%%%%%%%
\section{Conclusions}\label{sec:conclusions}

In this paper, we have presented methods and results for numerical CY metrics in the context of machine learning algorithms. Broadly, we have made progress in four directions. 1) The library \texttt{cymetric} presented here contains functions which can be applied to the numerical computation of CY metrics on a wide range of manifolds, namely to complete intersection CYs in products of projective spaces (CICYs) and to CY manifolds obtained from the Kreuzer-Skarke (KS) list. 2) In particular, we have realized point sampling for CY hypersurfaces in general toric ambient spaces. 3) We have shown how line bundle slopes can be used to obtain the numerical CY metric for given points in K\"ahler moduli space. 4) We have used the \texttt{cymetric}  library to obtain new results on the bi-cubic CY, including the computation of volumes, slopes and Hermitian Yang-Mills (HYM) connections on line bundles, and on a  Picard number two CY from the KS list.

The \texttt{cymetric} library consists of two main parts: the point generator and neural networks which are used to predict the CY metric.
For CICYs it is known for some time how to generate point samples with a defined distribution, following an algorithm explained in Refs.~\cite{Douglas:2006rr,Braun:2007sn} which is based on a theorem by Shiffman and Zelditch~\cite{Shiffman:1999aaa}. For CY hypersurfaces in toric varieties we have generalized this algorithm, making use of projective embeddings and the Shiffman-Zelditch theorem. Both versions of this algorithm, covering CICYs and KS manifolds, have been realized in the \texttt{cymetric} library.

The CY metric can be represented by a neural network in a number of ways and the \texttt{cymetric} library contains five possible realizations, as summarized in Table~\ref{tab:ansatz}. To ensure that all NNs converge to the Ricci-flat CY metric for the specified values of the moduli, we use a custom loss function with five contributions. They encode the Monge-Amp\`ere equation, the Ricci scalar, the K\"ahler property of the metric, the correct transition functions between coordinate patches, and the correct K\"ahler class. For the CY manifolds considered in this papers, the so called $\phi$-network (the last entry in Table~\ref{tab:ansatz}), built from an exact correction to a reference K\"ahler metric, performs the best. This is largely explained by the fact that this Ansatz guarantees the K\"ahler and transition loss conditions are automatically satisfied. However, other network Ans\"atze might well turn out to be useful for different CY manifolds. They can also be adapted to non-K\"ahler situations, for example in the context of special structure manifolds with torsion~\cite{Anderson:2020hux}.

In addition to testing our library on the quintic CY we have also performed runs on the bi-cubic CY, a manifold with Picard number two.  On this manifold, we have computed the numerical metric at seven specific points in K\"ahler moduli space. The results have been used to compute, via numerical integration with the CY metric, the CY volumes and some line bundle slopes at those seven points in moduli space. These quantities can also be computed exactly by intersection formulae and the good agreement of our numerical results, typically at the level of $\sim 1\%$ or smaller for the volume and a few percent for the slope, confirms that the numerical metrics have indeed been computed at the correct points in moduli space. We have also considered a Picard number two CY from the KS list and have carried out a computation of the CY metric at a specific point in its K\"ahler moduli. Again, we were able to reproduce the CY volume from numerical integration, indicating that the toric  point generator works correctly and that the metric is indeed computed at the desired point in K\"ahler moduli space.

In a final step we have used the numerical CY metric to find numerical solutions of the HYM equation for line bundles  on the bicubic. A necessary and sufficient condition for the existence of such solutions is that the line bundle has vanishing slope, a condition which can, depending on the choice of the bundle, be satisfied at some particular locus in K\"ahler moduli space. Using the  \texttt{cymetric} package, we have computed the numerical CY metric at such a point in moduli space. At this point we have then numerically solved the HYM equation for the line bundle metric, which can be cast in the form of an inhomogeneous Laplace equation. In this way, we find a numerical HYM line bundle metric at percent-level accuracy.

The goal is for \texttt{cymetric} to become an active open source community project similar to popular projects in the machine learning community. To this end we encourage researchers to include their own models and extensions. In the future, we plan to expand its capabilities by adding new models and neural network architectures based on more extensive ablation studies. It is important to emphasize that algebraic metrics such as in Ref.~\cite{Headrick:2009jz} can be implemented as special cases, by creating a custom network that just learns the coefficients in the polynomial approximating the K\"ahler potential. The same holds true for learning metrics, akin to the $h$-balanced metric from Donaldson's algorithm, on holomorphic sections of line bundles. The package then allows the full power of the TensorFlow optimizer to be applied in this context. We hope that the generality and flexibility of the package will help to solve important physics problems, such as, for example, the calculation of the physical Yukawa couplings from string theory.

There are a number of interesting additions, modifications and new directions which should be mentioned. It may be interesting to try out different methods for point sampling, for example a Markov Chain Monte Carlo sampling based on the measure $dV_\Omega$, in order to complement (and check) the methods based on the Shiffman and Zelditch theorem. The  \texttt{cymetric}  library allows computing the CY metric at given, specified points in moduli space. It would be interesting to use this feature and study the dependence of the CY metric on the moduli in more detail. We should also point out that, while the present paper has focused on CY three-folds, the \texttt{cymetric}  library can be applied to CY $n$-folds more generally. For example, it would be interesting to compute CY metrics on K3 and compare with recent analytic solutions found in Ref.~\cite{Kachru:2018van,Kachru:2020tat}, or compute a CY metric on a four-fold. Another direction is to investigate whether the present methods generalize to related settings, such as to manifolds with SU(3)-structure~\cite{Larfors:2018nce} or manifolds with $G_2$ holonomy. Work in this direction is in progress. 

\section*{Acknowledgments}
We thank Callum Brodie for useful discussions. The work of FR is supported by startup funding from Northeastern University. The work of ML and RS is funded by Vetenskapsr\aa det, under grant 2020-03230. Computations were in part enabled by resources provided by the Swedish National Infrastructure for Computing (SNIC) at the HPC cluster \emph{Tetralith}, partially funded by the Vetenskapsr\aa det through grant agreement no.\ 2018-05973, the hydra cluster of the Physics Department at the University of Oxford, and the computing cluster at CERN.

%%%%%%%%%%%%%%%%%%%%%%%%%%%%%%%%%%%%%%%%%%%%%%%%%%%%%%%%%%%%%%%%%%%%%%%%%%%%
\appendix
\section{Some toric geometry}
\label{app:ToricGeometryIntro}
We collect some results from toric geometry that are used in the main part of the text. Toric varieties serve as ambient spaces for the CY manifolds we want to study. We will restrict the discussion to the case where the CY manifold is given as a hypersurface in a toric ambient space or as a complete intersection in a direct product of projective ambient spaces. We will focus on the parts relevant to this setup only, and use succinct and simple exposition, somewhat at the expense of precise terminology (for example, we will be referring to finitely generated strongly convex rational polyhedral cones simply as cones). For an introduction to toric geometry see the excellent textbooks~\cite{Cox:2011aaa} and \cite{Fulton:1993aaa}.

Throughout the discussion, we focus on the more general toric case, and point out how this simplifies for (products of) projective spaces along the way.

\subsection{Fan and cones}\label{app:FanCones}
A $d$-dimensional toric variety is described in terms of a fan, which is a collection of $k$-dimensional cones, where $k=0,1,\ldots, d$. The one-dimensional cones $v_i$ are called rays, and they generate a lattice $N\subseteq \mathbbm{Z}^d$. Each vertex $v_i\in N$, $i=1,\ldots,n$ corresponds to one homogeneous coordinate $x_i$ of the toric ambient space. Their toric scalings $q_\alpha$ are given by the kernel of the vertex matrix:
\begin{align}
q_\alpha^i v_i = 0 \qquad \Rightarrow\qquad (x_1,x_2,\ldots,x_n)~\to~(\lambda_\alpha^{q_\alpha^1}x_1,\lambda_\alpha^{q_\alpha^2}x_2,\ldots,\lambda_\alpha^{q_\alpha^n}x_n)\,.
\end{align}
The matrix $q_\alpha^i$ is called the GLSM charge matrix or the Mori cone matrix. Note that for a given set of charge vectors $\vec{q}_\alpha$, we can form a new set $\vec{q}_\beta=R^{\alpha}_{\beta}\vec{q}_\alpha$ by rotating the charges. One typically chooses the basis such that the GLSM D-terms
\begin{align}
\sum_i q_\alpha^i |x_i|^2 = t_\alpha
\end{align}
have a solution for K\"ahler parameters $t_\alpha\geq 0$. 

For a projective ambient space $\bP^{n_\alpha}$, the GLSM charge matrix is simply $q_\alpha^i=(1,1,\ldots,1)$ with $i=1,\ldots,n_\alpha+1$. Hence, for products of projective ambient spaces, we obtain a block-diagonal charge matrix . In contrast to the toric case, this does not encode any information beyond what is specified by just giving the dimensions $n_\alpha$ of the projective ambient space factors.

The top-dimensional, that is, $d$-dimensional, cones are called facets. All other cones are typically just referred to as faces. A $k$-dimensional cone $\sigma_{i_1,i_2,\ldots,i_k}$ in $N_\mathbbm{R}=N\times\mathbbm{R}$ is generated by vertices $v_{i_1}, v_{i_2}, \ldots, v_{i_k}$,
\begin{align}
\sigma_{i_1,i_2,\ldots,i_k} = \left\{\left.\sum_{i=1}^k\lambda_iv_i\in N_{\mathbbm{R}}~\right|~\lambda_i\geq0\quad\forall k \right\}\,.
\end{align}
Associated to this, we can define a dual cone $\sigma^*$ in the dual lattice $M_\mathbbm{R}\subseteq\mathbbm{Z}^d$,
\begin{align}
\label{eq:dualcones}
\sigma^*_{i_1,i_2,\ldots,i_k} = \left\{\left. w\in M~\right|~\langle v,w\rangle\geq0\quad\forall v\in\sigma_{i_1,i_2,\ldots,i_k}\right\}\,.
\end{align}
Note that the dual of a top-dimensional cone $\sigma_{i_1,i_2,\ldots,i_d}$ is again $d$-dimensional.

\subsection{Affine patches}\label{app:AffinePatches}
Each top-dimensional cone $\sigma$ of a fan corresponds to an affine patch $\mathbbm{C}^d$. The coordinates that are allowed to vanish in this patch are the ones given by the generators of $\sigma$, while all others are non-vanishing. We can construct the affine coordinates $z_\mu$ from the dual cone $\sigma^*$ by taking the product of all coordinates $x_i$ with exponents given by the inner product of the corresponding vertex $v_i$ with the generators $w_\mu$ of the dual cone $\sigma^*$, so that
\begin{align}
\label{eq:affinecoordinates}
z_\mu=\prod_{i=1}^n x_i^{\langle v_i,w_\mu \rangle}\,.
\end{align}
Note that in particular $\sigma^*$ is $d$-dimensional. Hence a toric variety of complex dimension $d$ always has $d$ affine coordinates in any given patch.  For a projective ambient space $\bP^{n_\alpha}$, this construction reduces to the standard patches $\mathcal{U}_a=\{z_a\neq0\}$ with affine coordinates $z_\mu=x_i/x_a$ for $i\neq a$.

\subsection{Divisors and Stanley-Reisner ideal}
Each vertex $v_i\in N$ corresponds to one homogeneous coordinate $x_i$ of the toric ambient space. Hence, we can introduce a set of (in general linearly dependent) divisors $D_i:=\{x_i=0\}$. Two ambient space divisors $D=\sum_i c_i D_i$ and $D'=\sum_i c_i' D_i$ for some $c_i,c'_i\in\mathbbm{Z}$ are linearly equivalent if all their GLSM charges agree,
\begin{align}
D\sim D' \qquad\Leftrightarrow\qquad \sum_i c_i q_\alpha^i = \sum_i c_i' q_\alpha^j\quad\forall\alpha\,.
\end{align}
In particular, this means that favorable CYs (that is, CYs which inherit the entire second cohomology from the ambient space) lead to the index range $\alpha=1,2,\ldots,h^{1,1}(X)$.

If certain divisors are not contained in a common cone of the fan, they cannot vanish simultaneously. We associate a monomial $s=\prod x_i$ to each such combination. The collection of these monomials generates an ideal,  
\begin{align}
\text{SRI}=\langle s_i \rangle\,,
\end{align}
called the Stanley-Reisner ideal. A toric variety is specified by the GLSM charges and the SRI, or, equivalently, by its vertices plus a chosen triangulation. 

For a projective space $\bP^{n_\alpha}$, the Stanley-Reisner ideal is generated by a single monomial, $\text{SRI}=\langle x_0x_1\ldots x_{n_\alpha}\rangle$, which corresponds to the familiar statement that not all coordinates of $\bP^{n_\alpha}$ can vanish simultaneously.

\subsection{Line bundles and sections}\label{app:Sections}
\label{sec:LineBundlesSections}
The sections of a line bundle $L$ associated to a divisor can be readily computed from the vertices $v_i$ of $N$ and their dual vertices $w$ in the dual lattice $M$:
\begin{align}
\label{eq:divisorsections}
D=\sum_{i=1}^n c_i D_i \quad\rightarrow\quad \Gamma(D) \sim\left\{\left. \prod_{i=1}^n x_i^{\langle v_i,w\rangle+c_i}~\right|~\langle v_i,w \rangle +c_i\geq 0\,,\quad w\in M \right\}\,.
\end{align}
Note that for non-negative $c_i$, only finitely many $w\in M$ satisfy the condition $\langle v_i,w \rangle +c_i\geq 0$. For a divisor $D$ inside a product of $n$ projective ambient spaces, this reduces to the statement that a section basis for $D$ is given by the monomials of multi-degree $(c_1,c_2,\ldots,c_n)$.

One can prove a theorem which will be important for constructing K\"ahler forms later. It states that for any nef divisor (or rather the sections of the associated nef line bundle), precisely one of the monomials that form a basis of the sections vanishes nowhere in a given patch. This means that the base-point free divisors are given by sections of nef line bundles.

Since the anti-canonical divisor of a toric variety is given by the sum over all divisors, $-K=\sum_{i=1}^n D_i$ where $n$ is the number of vertices, this motivates defining a lattice polytope $P\in N$ given by the convex hull of all vertices $v_i$, and its dual lattice polytope  $P^*\in M$ in the dual $M$-lattice via
\begin{align}
P^*:=\left.\left\{w_i\in N~\right|~\langle v_i,w_j\rangle\geq-1\quad\forall v_i\in P\right\}\,.
\end{align}
For the purpose of CYs given as hypersurfaces in toric varieties, we only need to look at reflexive polytopes, which are lattice polytopes whose dual are again lattice polytopes. All reflexive polytopes have the origin as their unique interior point. Their dual is also a reflexive polytope and satisfies $(P^*)^*=P$. With this, it is easy to find a basis of sections of the anti-canonical bundle and which then allows writing down the most general CY defining equation
\begin{align}
\label{eq:hypersurface}
p=\sum_{w\in P^*}c_w \prod_{i=1}^n x_i^{\langle v_i,w\rangle}\,.
\end{align}
Here, the coefficients $c_w\in\bC$ parameterize the complex structure of the CY.

In the case of ambient spaces $\mathcal{A}$ that are a product of projective spaces, $\mathcal{A}=\otimes_{\alpha=1}^N\bP^{n_\alpha}$, we are considering complete intersections of $k$ hypersurfaces $H_i$, $i=1,\ldots, k$ rather than just a single hypersurface. This requires specifying the multi-degrees, $c^\alpha_i$, of the normal bundle sections whose complete intersection defines the CY manifold. The CY condition amounts to the constraints $\sum_i c^\alpha_i=n_\alpha+1$, where $\alpha=1,\ldots,k$, on these multi-degrees. (In the toric case, this gives rise to nef partitions of the reflexive polytope, but we will not consider such cases here, even though our methods still apply.) Moreover, for a CY $d$-fold, we require that $\sum_\alpha n_\alpha - k = d$. This means that we can specify a complete intersection CY manifold inside an ambient space that is given as a product of projective spaces purely by specifying the degrees $c^\alpha_i$, which form a matrix referred to as the 'configuration matrix' of the CICY. 

%%%%%%%%%%%%%%%%%%%%%%%%%%%%%%%%%%%%%%%%%%%%%%%%%%%%%%%%%%%%%%%%%%%%%%%%%%%%

\bibliographystyle{bibstyle}
\bibliography{refs}
\end{document}